\documentclass[10pt,aps,prx,twocolumn]{revtex4-2}

\usepackage{graphicx}
\usepackage{amssymb}
\usepackage{amsmath}
\usepackage{xcolor}
\usepackage{hyperref}
\hypersetup{
    colorlinks=true,
    linkcolor=blue,
    citecolor=blue,
    filecolor=blue,      
    urlcolor=blue,
}

\DeclareMathOperator{\sech}{sech}
\usepackage{comment}

\begin{document}

\title{Self-regulating soliton domain walls in microresonators}
\author{Heming Wang$^{1,\dag}$, Boqiang Shen$^{1,\dag}$, Yan Yu$^1$, Zhiquan Yuan$^1$, Chengying Bao$^1$, Warren Jin$^2$, Lin Chang$^2$, Mark A. Leal$^2$, Avi Feshali$^3$, Mario Paniccia$^3$, John E. Bowers$^{2,*}$, and Kerry Vahala$^{1,*}$\\
$^1$T. J. Watson Laboratory of Applied Physics, California Institute of Technology, Pasadena, CA 91125, USA\\
$^2$ECE Department, University of California Santa Barbara, Santa Barbara, CA 93106, USA\\
$^3$Anello Photonics, Santa Clara, CA\\
$^*$jbowers@ucsb.edu, vahala@caltech.edu}

\begin{abstract} 
Dissipative soliton Kerr frequency combs in microresonators have recently been demonstrated with the self-injection locking process. They have the advantage of turnkey deterministic comb generation and simplifying dark soliton generation in the normal dispersion regime. Here, the formation process of dark pulses triggered by self-injection locking is studied by regarding them as a pair of domain walls that connect domains having different intracavity powers. The self-injection locking mechanism allows the domain walls to self-regulate their position so that a wide range of dark comb states can be accessed, and the duty cycle is controlled by the feedback phase. Direct imaging of the dark pulse shape using the electro-optic sampling technique is used to verify the theory. The results provide new physical insights as well as a new operational modality for this important class of nonlinear waves.
\end{abstract}

\maketitle

\section{Introduction}

Soliton microcombs \cite{kippenberg2018dissipative} offer a path towards miniaturization of optical frequency comb technologies \cite{diddams2020optical} onto photonic chips. Their integration with III-V pump lasers without the need for optical isolation \cite{stern2018battery, raja2019electrically, shen2020integrated,jin2021hertz} is an important step towards fully integrated chip-based soliton microcombs. The self-injection locking process \cite{razavi2004study}, which was originally used to reduce laser linewidth \cite{liang2010whispering, liang2015high, kondratiev2017self}, has been shown to create a new ``turnkey'' operating point \cite{shen2020integrated} that eliminates complex startup and feedback protocols \cite{voloshin2021dynamics}. This combination of features enables single-chip soliton microcomb devices that comprise heterogeneously integrated III-V/Si pump lasers and microresonators \cite{xiang2021laser}. Moreover, the self-injection-locked ``turnkey'' operation simplifies access to dark pulse states \cite{jin2021hertz, lihachev2021platicon}. Specifically, dark pulses exist under conditions of normal group velocity dispersion (GVD) \cite{liang2014generation, xue2015mode, huang2015mode, lobanov2015frequency, xue2017microresonator} and their formation normally requires special spectral-design considerations (e.g., mode-crossing induced anomalous dispersion \cite{xue2015mode}). Self-injection locking makes it possible to turnkey-trigger dark pulses without these requirements, and instead relies on the intrinsic Rayleigh backscattering ubiquitous in resonators \cite{jin2021hertz, lihachev2021platicon}. However, despite this benefit, a theory describing the normal dispersion microcomb generation process under conditions of self injection locking has not yet been established. 

Here we analyze the formation process of such dark pulses in the self-injection locking regime. It is shown that nonlinear injection locking not only eliminates the startup protocols from a technical viewpoint \cite{shen2020integrated, jin2021hertz}, but also provides a new physical understanding of these pulses, wherein two oppositely-oriented domain walls are able to regulate their own dynamics. Moreover, the set point in this self-regulation is controlled by the feedback phase so that the duty cycle can be adjusted to vary comb spectra and optimize comb power efficiency. In optics, polarization domain walls and novel types of vector dark domain wall solitons have been theoretically predicted \cite{haelterman1994polarization} and observed in a fiber ring laser \cite{zhang2009observation, zhang2010vector}, and we will reveal the similarities between the structure studied here and the previous optical domain walls, including the existence of an exchange symmetry. In the context of resonators, the concept of switching waves \cite{rozanov1982transverse} has also been introduced. Their dynamics are governed by energy balance and can be described by the Maxwell point \cite{parra2016origin}, which plays a central role in pulse formation and self-injection feedback. A model is developed and validated by taking ``snapshots'' of dark pulse shapes via the electro-optic sampling technique \cite{ferdous2009dual, duran2015ultrafast, yi2018imaging}.

This paper is organized as follows.
In section II we begin with a model for nonlinear injection locking and arrive at the Lugiato-Lefever equation (LLE) augmented with a locking condition. Some general properties of these equations are also summarized here.
In Section III we introduce the mechanism for domain and domain wall formation by considering the zero dispersion case, and use these results to demonstrate the physical idea behind domain wall generation.
In Section IV we move on to the case of normal dispersion, where the energy balance of the domain wall leads to the concept of Maxwell point.
In section V we demonstrate how the domain walls self-regulate around the Maxwell point.
In Section VI the effects of feedback phase on pulse numbers and duty cycle are studied, and the dependence is utilized for comb efficiency calculations.
In Section VII we present some preliminary experimental results validating the model.
Finally in Section VIII we discuss possible improvements to the model.
Various technical derivations are collected in the appendixes.

\section{The nonlinear injection locking model}

\begin{figure*}
\includegraphics[width=170mm]{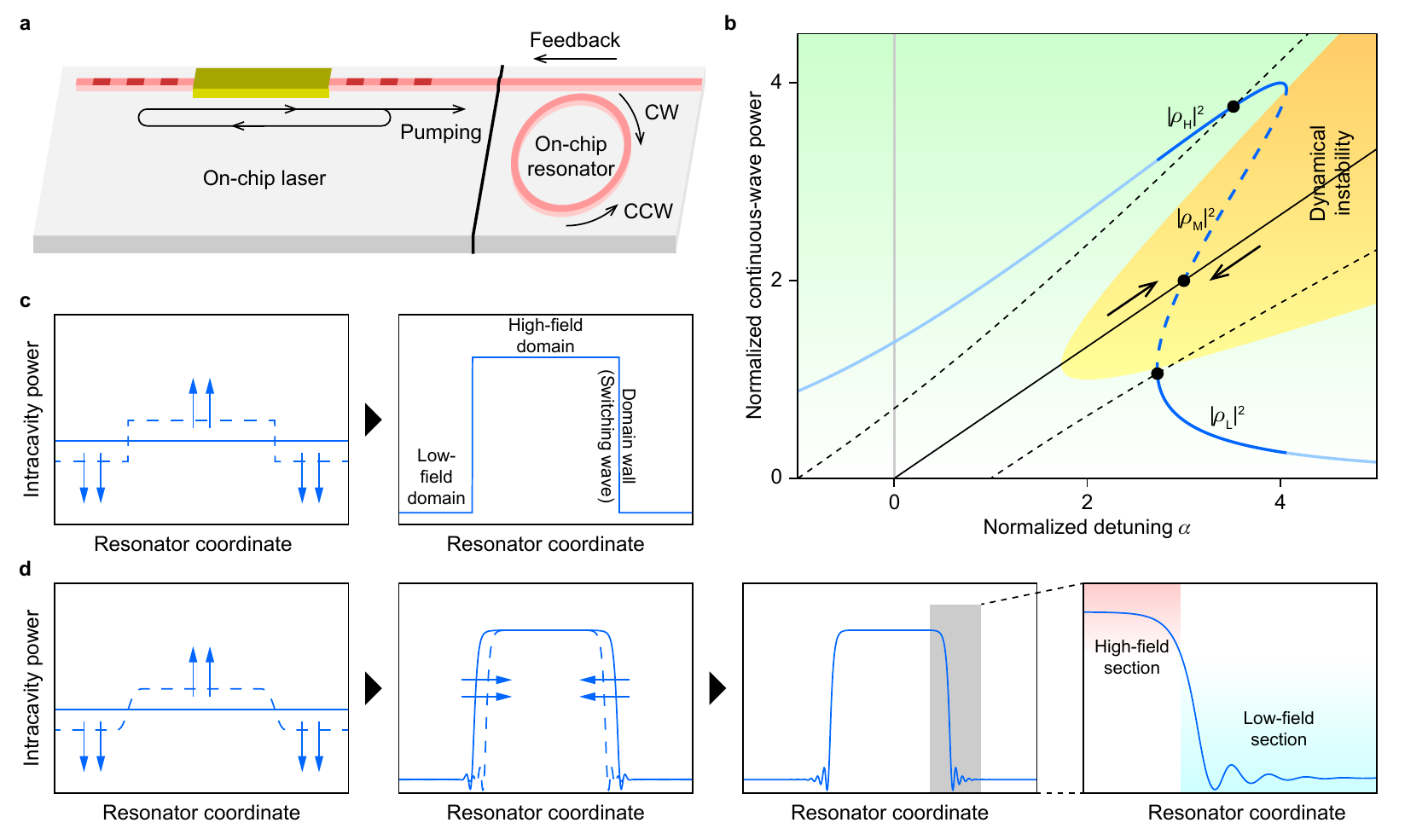}
\caption{\label{FIG1}
{\bf Laser-resonator system with nonlinear injection locking.}
(a) Schematic of the system, where an on-chip laser is coupled to an on-chip microresonator without optical isolation, thereby allowing signals from the resonator to be fed back to the laser. Within the resonator, light circulates in both the clockwise (CW) and counter-clockwise (CCW) directions.
(b) Blue curve shows the nonlinear resonator pumping curve for $|f|^2=4$, and the dashed blue line marks the section that is dynamically unstable. The orange region (marked with ``Dynamical instability'') gives the DI region when the pump power varies. The bold blue sections give the three branches of the multivalued part of the curve, and the powers correspond to $|\rho_\mathrm{H}|^2$, $|\rho_\mathrm{M}|^2$ and $|\rho_\mathrm{L}|^2$. Black lines show the laser locking curve in the presence of injection locking. The feedback phases are taken as $-\pi/2$, $0$ and $\pi/2$ (from left to right), where the $\phi=0$ curve is solid and others are dashed. Arrows show the evolution direction of the system. The black dots mark continuous-wave operating points of the system associated with these phases. 
(c) Schematic of intracavity field evolution in the absence of dispersion. Left panel: after the intracavity field reaches $\rho_\mathrm{M}$ at the continuous-wave operating point, fluctuations of the field causes the field evolving towards the upper and lower equilibria. Arrows show the evolution direction of the respective fields. Right panel: high- and low-field domains appear in the resonator, and a domain wall forms to connect the two domains.
(d) Schematic of intracavity field evolution in the presence of normal dispersion. First panel: similar to the dispersionless case, the field evolves towards the upper and lower equilibria after reaching the continuous-wave operating point. Second panel: after domain walls form in the resonator, the walls adjust their position through the regulation process. Third panel: the field reaches steady state at the Maxwell point. Fourth panel: the region marked in the third panel is enlarged, showing the domain wall solution. The red and blue areas mark the high-field and low-field domains, respectively.}
\end{figure*}

We consider a self-injection system consisting of a nonlinear ring-type resonator and a laser as shown in Fig. \ref{FIG1}a. The laser and resonator are directly coupled without optical isolation, allowing the backscattered light from the resonator to be fed back to the laser and alter its dynamics. For the forward field in the resonator, its equation of motion reads:
\begin{align}
\frac{\partial E_\mathrm{F}}{\partial t}
&=-\left(\frac{\kappa}{2}+i\delta\omega\right)E_\mathrm{F}
+i\frac{D_2}{2}\frac{\partial^2 E_\mathrm{F}}{\partial \theta^2}\nonumber\\
&+ig_\mathrm{NL}\left(|E_\mathrm{F}|^2+2\int_0^{2\pi}|E_\mathrm{B}|^2\frac{d\theta}{2\pi}\right)E_\mathrm{F}\nonumber\\
&+ig_\mathrm{L}^*\star E_\mathrm{B}
+i\sqrt{\kappa_\mathrm{ex}}F_\mathrm{in},
\end{align}
where $E_\mathrm{F}$ ($E_\mathrm{B}$) is the forward (backward) slowly-varying field amplitude normalized to energy, $\kappa$ is the energy loss rate for the modes (assumed to be the same for each spectral mode), $\delta\omega$ is the instantaneous detuning between the cold resonance being pumped and the laser, $D_2$ is the second-order dispersion parameter, $g_\mathrm{NL}$ is the nonlinear coefficient, $g_\mathrm{L}$ is the distributed linear scattering strength, $\kappa_\mathrm{ex}$ is the external coupling rate to the waveguide, $F_\mathrm{in}$ is the input amplitude on the waveguide normalized to power, $\theta$ is the co-moving resonator coordinate (proportional to the fast time) and $t$ is the slow time. The nonlinearity from $|E_\mathrm{F}|$ is localized, while the nonlinearity from $|E_\mathrm{B}|$ is averaged over the entire cavity as these fields propagate in opposite directions and do not phase match with $E_\mathrm{F}$. The convolution form of the linear scattering, $[g_\mathrm{L}^*\star E_\mathrm{B}](\theta)=\int_0^{2\pi}g_\mathrm{L}^*(\theta')E_\mathrm{B}(\theta-\theta')d\theta'$ represents general elastic scattering that may contain both continuous (e.g. surface roughness) and discrete sources (e.g. individual particles). The equation here only includes those effects that are necessary for soliton generation, i.e. detuning, dispersion, nonlinearity, loss and pumping. Other effects can be readily accommodated, such as high-order dispersion, Raman effects, and different losses on each spectral mode, by adding or modifying the corresponding terms in the equation. A similar equation holds for the backward amplitude $E_\mathrm{B}$ except there is no external pumping term.

The laser dynamics includes both gain and loss, and reads:
\begin{align}
\frac{\partial E_\mathrm{L}}{\partial t}
&=i\left(\delta\omega_\mathrm{L}-\delta\omega\right)E_\mathrm{L}
+(1+i\alpha_\mathrm{G})\left(-\frac{\gamma}{2}+G\right)E_\mathrm{L}\nonumber\\
&+i\sqrt{\gamma}F_\mathrm{L,in},
\end{align}
where $E_\mathrm{L}$ is the slow-varying amplitude in the laser cavity, $\delta\omega_\mathrm{L}$ is the detuning of the cold resonance compared to the free-running laser, $G$ is the laser gain that depends on $|E_\mathrm{L}|^2$ through gain saturation, $\gamma$ is the laser cavity loss, $\alpha_\mathrm{G}$ is the amplitude-phase coupling factor, and we replaced out-coupling loss with $\gamma$ by assuming that this is the dominant loss source of the cavity. Unlike $E_\mathrm{F}$ and $E_\mathrm{B}$, $E_\mathrm{L}$ can be treated as a complex number rather than a spatially-dependent field. This is possible because the pump laser used is single mode and also because the dominant source of resonator feedback is considered to be from backscattering of the pump wave. Carrier dynamics are also ignored as these dynamics are generally much faster than the time scale of power change (i.e, $1/\kappa$) associated with the pumping field in the high-Q cavity. Typical carrier relaxation rates for semiconductor lasers can be as large as a several GHz while state-of-the-art integrated resonators reach a resonance linewidth of a MHz or less.

The external (in the waveguide) pumping for the resonator and laser are related to the internal fields through the input-output relations,
\begin{align}
F_\mathrm{in}=i\sqrt{\gamma}\sqrt{T}\exp(i\phi_\mathrm{B})E_\mathrm{L},\\
F_\mathrm{L,in}=i\sqrt{\kappa_\mathrm{ex}}\sqrt{T}\exp(i\phi_\mathrm{B})\overline{E_\mathrm{B}},
\end{align}
where $T$ is the power transmission on the feedback waveguide, including all waveguide loss and facet coupling loss accumulated along the waveguide. $\phi_\mathrm{B}$ is the phase accumulated on the feedback waveguide, and $\overline{E_\mathrm{B}}=(2\pi)^{-1}\int_0^{2\pi}E_\mathrm{B}d\theta$ is the average field amplitude (the amplitude on the zeroth mode) for $E_\mathrm{B}$. We assume that $\phi_\mathrm{B}$ is a constant over the bandwidth being considered, which requires that the feedback length is short. The approximation of using the spatial average of the backscattered field is partially justified because the pumping field intensity is typically larger than that of all comb lines and the single-mode laser resonator will tend to reject inputs at other frequencies (i.e., they are non-resonant).

For each of the above equations, we do not require $\delta\omega$ to be a constant over time as the laser frequency can shift around while tracking the resonance. The equations are always referenced to the instantaneous detuning of the laser, such that the pumping term no longer contains any explicit frequency terms. In effect, this ensures that the pumping term can be taken as a positive real number for later convenience.

The above equations for resonators and lasers, while useful in numerical simulations, are not suitable for studying the dynamics from a theoretical perspective. To simplify the model and reveal the underlying physics, some necessary approximations are made such that terms that do not contribute significantly to the pulse formation and stabilization process are discarded. Relaxation of some of these approximations is addressed in the discussion section. The principal assumption used here is that the backscattering is weak, i.e. $g_\mathrm{L}\ll\kappa$, which is often the case in current experiments. As frequency shifts caused by nonlinear effects are also on the order of $\kappa$ when the comb forms, this allows us to drop nonlinearity terms induced by $|E_\mathrm{B}|^2$. We will also neglect the $g_\mathrm{L}^*\star E_\mathrm{B}$ term in $\partial_t E_\mathrm{F}$ that scales as $|g_\mathrm{L}|^2$ in this weak-scattering approximation. As a result, all mode amplitudes of $E_\mathrm{B}$ will be decoupled from the system except the zeroth mode, which is determined by the pumping field, and we can replace the field $E_\mathrm{B}$ by its zeroth mode amplitude $\overline{E_\mathrm{B}}$ (the inclusion of the comb lines will be numerically considered below). The resonator equations then simplify to:
\begin{align}
\frac{\partial E_\mathrm{F}}{\partial t}
&=-\left(\frac{\kappa}{2}+i\delta\omega\right)E_\mathrm{F}
+i\frac{D_2}{2}\frac{\partial^2 E_\mathrm{F}}{\partial \theta^2}\nonumber\\
&+ig_\mathrm{NL}|E_\mathrm{F}|^2E_\mathrm{F}
-\sqrt{\kappa_\mathrm{ex}\gamma T}E_\mathrm{L}\exp(i\phi_\mathrm{B}),
\end{align}
\begin{align}
\frac{d \overline{E_\mathrm{B}}}{d t}
&=-\left(\frac{\kappa}{2}+i\delta\omega\right)\overline{E_\mathrm{B}}
+2ig_\mathrm{NL}\overline{E_\mathrm{B}}\int_0^{2\pi}|E_\mathrm{F}|^2\frac{d\theta}{2\pi}\nonumber\\
&+i\overline{g_\mathrm{L}}\overline{E_\mathrm{F}},
\end{align}
where $\overline{g_\mathrm{L}}=\int_0^{2\pi}g_\mathrm{L}(\theta')d\theta'$ is the backscattering strength for the zeroth mode.

The laser dynamics for $E_\mathrm{L}\equiv |E_\mathrm{L}|\exp(i\phi_\mathrm{L})$ can be split into amplitude and phase parts:
\begin{equation}
\frac{1}{|E_\mathrm{L}|}\frac{d|E_\mathrm{L}|}{dt}  =-\frac{\gamma}{2}+G-\mathrm{Re}\left[\sqrt{\kappa_\mathrm{ex}\gamma T}e^{i\phi_\mathrm{B}}\frac{\overline{E_\mathrm{B}}}{E_\mathrm{L}}\right]
\label{Eq_ELamp}
\end{equation}
\begin{equation}
\frac{d\phi_\mathrm{L}}{dt} = \delta\omega_\mathrm{L}-\delta\omega+\left(-\frac{\gamma}{2}+G\right)\alpha_g-\mathrm{Im}\left[\sqrt{\kappa_\mathrm{ex}\gamma T}e^{i\phi_\mathrm{B}}\frac{\overline{E_\mathrm{B}}}{E_\mathrm{L}}\right]
\label{Eq_ELphase}
\end{equation}
In accordance with the earlier discussion, we assume that the laser relaxation dynamics are fast enough such that the laser power adiabatically tracks the external input from backscattering ($d|E_\mathrm{L}|/dt\approx 0$). With these assumptions, the instantaneous gain can be solved from the amplitude equation (\ref{Eq_ELamp}) and eliminated from the phase equation (\ref{Eq_ELphase}). This results in
\begin{align}
0&=\delta\omega_\mathrm{L}-\delta\omega\nonumber\\
&-\mathrm{Im}\left[(1-i\alpha_\mathrm{G})\sqrt{\kappa_\mathrm{ex}\gamma T}\exp(i\phi_\mathrm{B})\frac{\overline{E_\mathrm{B}}}{E_\mathrm{L}}\right],
\end{align}
so that $E_\mathrm{L}$ is now reduced from a dynamical variable to a parameter (i.e. the laser power is almost unchanging).

We now normalize all variables in the equations. The normalization scheme is based on $\kappa/2\rightarrow 1$ and $g_\mathrm{NL}\rightarrow 1$. Phase changes in different variables are also merged together.
Define the normalized detunings $\alpha=2\delta\omega/\kappa$ and $\alpha_\mathrm{L}=2\delta\omega_\mathrm{L}/\kappa$,
normalized time $\tau=\kappa t/2$,
normalized field $\psi=E_\mathrm{F}\sqrt{2g_\mathrm{NL}/\kappa}$,
normalized average $\rho=\overline{E_\mathrm{F}}\sqrt{2g_\mathrm{NL}/\kappa}$ and $\rho_\mathrm{B}=\overline{E_\mathrm{B}}\sqrt{2g_\mathrm{NL}/\kappa}$,
normalized dispersion $\beta_2=-2D_2/\kappa$ (the negative sign here follows the sign convention for the group velocity dispersion, GVD),
normalized backscattering $\beta=2\overline{g_\mathrm{L}}/\kappa$,
normalized pump $f=-(2/\kappa)^{3/2}\sqrt{g_\mathrm{NL}\kappa_\mathrm{ex}\gamma T}E_\mathrm{L}\exp(i\phi_\mathrm{B})$ (we will take $f$ as a positive real number without loss of generality from here on),
and average power for the forward mode $P=\int_0^{2\pi}|\psi|^2d\theta/(2\pi)$.
After normalizing all the variables, we arrive at the following set of equations:
\begin{align}
\frac{\partial \psi}{\partial \tau} & = -(1+i\alpha)\psi-i\frac{\beta_2}{2}\frac{\partial^2 \psi}{\partial\theta^2}+i|\psi|^2\psi+f\\
\frac{d \rho_\mathrm{B}}{d \tau} & = -(1+i\alpha-2iP)\rho_\mathrm{B}+i\beta\rho\\
\alpha & = \alpha_\mathrm{L}+K\mathrm{Im}\left[e^{i\phi}\frac{\rho_\mathrm{B}}{i\beta f}\right]
\end{align}
where we introduced two additional parameters: the (normalized) locking bandwidth,
\begin{equation}
K=\frac{4\kappa_\mathrm{ex}\gamma}{\kappa^2}\sqrt{1+\alpha_\mathrm{G}^2}|\beta|T
\end{equation}
and the feedback phase,
\begin{equation}
\phi=2\phi_\mathrm{B}+\mathrm{Arg}[\beta]-\mathrm{arctan}(\alpha_\mathrm{G})+\frac{\pi}{2}
\end{equation}
where $\mathrm{Arg}[\cdot]$ is the argument function. The feedback phase $\phi$ consists of three parts: optical phase accumulated on the waveguide, backscattering, and amplitude-phase coupling. The extra $\pi/2$ is added to the definition of $\phi$ for later convenience. The first equation is identical to the normalized Lugiato-Lefever equation (LLE) as we have neglected all terms that do not contribute significantly to the comb formation process. The second and last equations resemble the Lang-Kobayashi equation \cite{lang1980external} and augment the LLE to describe the nonlinear self-injection locking process.

For the following analyses we will work with the limiting case that $K\rightarrow\infty$, such that the locking process completely overrides the free-running laser detuning $\alpha_\mathrm{L}$, and the laser is always locked to the detuning determined by the implicit equation $\mathrm{Im}[e^{i\phi}\rho_\mathrm{B}/(i\beta f)]=0$. This can be justified as most semiconductor laser resonators possess a much lower $Q$ (typically $10^4$ to $10^5$) compared to that of the resonator used for comb generation (typically around $10^{8}$), and $K$ can reach $10^2$ even with relatively weak backscattering ($\beta \approx 10^{-2}$). If we further assume steady-state conditions for the backward field $\rho_\mathrm{B}$, the locking condition can be expressed as
\begin{equation}
\mathrm{Im}\left[\frac{e^{i\phi}}{1+i\alpha-2iP}\frac{\rho}{f}\right]=0.
\label{Eq_locking}
\end{equation}
This will be referred to as the ``locking curve'' equation. For the resonator, the steady-state continuous-wave power under external pumping can be found through
\begin{equation}
f^2=[1+(\alpha-|\rho|^2)^2]|\rho|^2,
\label{Eq_pumping}
\end{equation}
and referred to as the ``pumping curve'' equation, which is plotted together with the locking curves with different feedback phases in Fig. \ref{FIG1}b. The resonator pumping curve may have three branches with respect to the detuning $\alpha$, and the field solutions are denoted as $\rho_\mathrm{H}$, $\rho_\mathrm{M}$ and $\rho_\mathrm{L}$, ordered by their absolute value from highest to lowest. Solutions on the upper ($\rho_\mathrm{H}$) and lower ($\rho_\mathrm{L}$) branches are readily shown to be stable while the middle branch solution ($\rho_\mathrm{M}$) is dynamically unstable under homogeneous perturbations in the temporal domain in the absence of injection locking. The dynamical instability (DI) region is indicated in Fig. \ref{FIG1}b, the boundaries of which can be found through $\partial\alpha/\partial|\rho|^2=0$ and solved as \cite{godey2014stability}
\begin{equation}
|\rho|^2=\frac{2\alpha}{3}\pm\frac{\sqrt{\alpha^2-3}}{3},\ \ \alpha\geq\sqrt{3}
\end{equation}
We note that the DI region marks the existence of optical bistability which enables the formation of dark pulses consisting of $\rho_\mathrm{H}$ and $\rho_\mathrm{L}$ continuous-wave components, and plays a special role in dark-pulse generation in the injection-locking scheme. It also falls within the modulational instability (MI) (which is unstable under inhomogeneous perturbations, a prerequisite of comb generation) region, which explains the onset of dark pulse generation from a frequency-domain perspective. The connection between DI and MI is further explored in Section VI.

For continuous-wave conditions, we can replace $\psi$ with $\rho$ and $P$ with $|\rho|^2$. The locking then results in
\begin{equation}
\mathrm{Im}\left[\frac{e^{i\phi}}{(1+i\alpha-i|\rho|^2)(1+i\alpha-2i|\rho|^2)}\right]=0.
\end{equation}
This is a quadratic equation in $\alpha$ and can be solved as
\begin{equation}
\alpha=\frac{3}{2}|\rho|^2-\cot\phi+\frac{\sqrt{4+|\rho|^4\sin^2\phi}}{2\sin\phi}
\label{Eq_cw_locking}
\end{equation}
with the understanding that for the case $\phi=0$, a limit of the above relation should be taken and results in $\alpha=3|\rho|^2/2$. This equation will be referred to as the “continuous-wave locking curve” equation, which describes the laser locking characteristics prior to comb generation as a function of $|\rho|^2$. A simple stability analysis shows that this root describes stable locking (the other root pushes the system away from the equilibrium). The continuous-wave locking curve [Eq. (\ref{Eq_cw_locking})] intersects the resonator pumping curve [Eq. (\ref{Eq_pumping})] exactly once at the ``continuous-wave operating point" under all $\phi$ and $f$, eliminating the possibility of multiple continuous-wave steady states in the system. If the continuous-wave operating point falls within the DI region, the injection locking process also makes the system dynamically stable (but still modulationally unstable). In this case, when pumping an initially unpumped resonator, the system is quickly pulled to the continuous-wave operating point on the middle unstable branch, after which dark pulses are generated (Fig. \ref{FIG1}c and \ref{FIG1}d), causing the system state to move further along the general laser locking curve [Eq. (\ref{Eq_locking})]. The detailed injection locking dynamics accompanied with dark pulse generation will be thoroughly discussed in Section III and IV. 

We note that there are out-of-lock states for the resonator if the laser is tuned sufficiently far away from the resonance. However, these do not show up in the current analysis as we are working with the $K\rightarrow\infty$ limit, and mainly concerned with the system behaviour within the mode rather than the locking bandwidth.

\section{Nonlinear injection locking with zero dispersion}

\begin{figure}
\includegraphics[width=85mm]{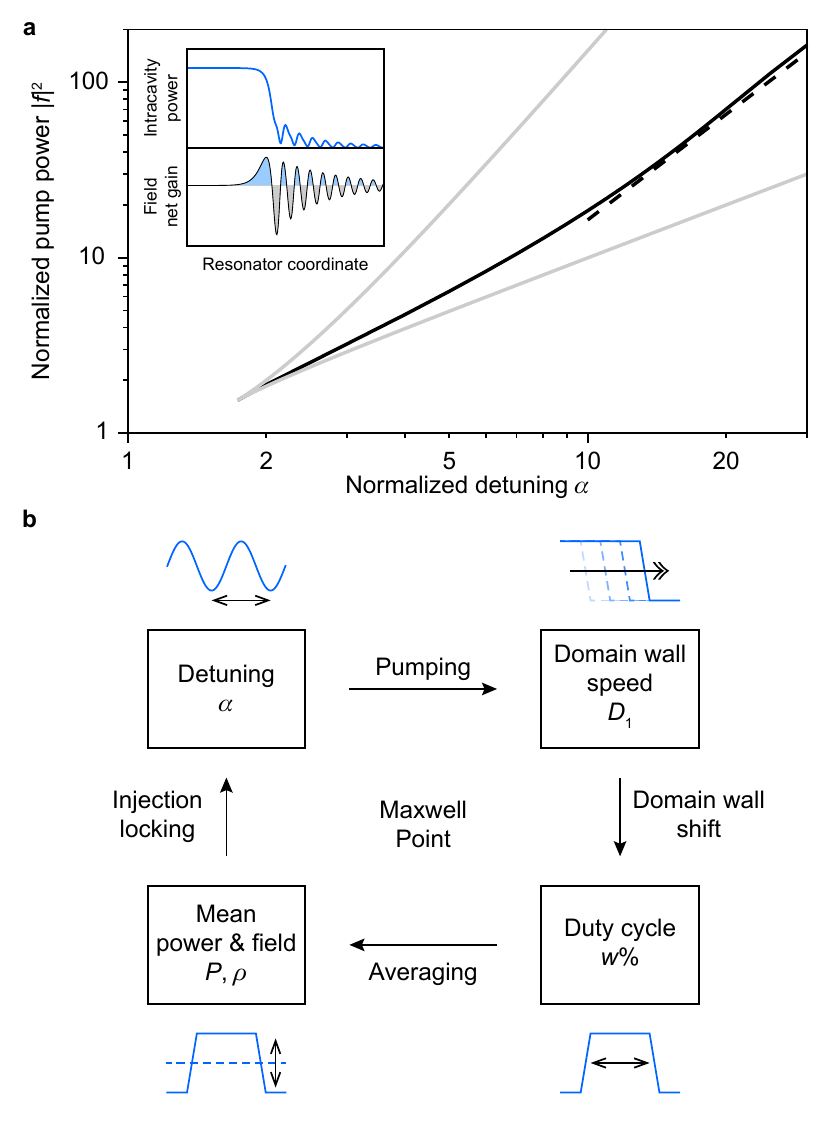}
\caption{\label{FIG2}
{\bf Maxwell point and the self-regulation mechanism.}
(a) The Maxwell point (black solid curve) as a function of detuning. Gray curves mark the DI region, which is approximate boundary of dark pulse generation. Eq. (\ref{Eq_MPVar}) is plotted as the black dashed line. Inset: top panel shows the domain wall solution for $\alpha=15$, and bottom panel shows the corresponding net field gain calculated as $-2|\psi|^2+2\mathrm{Re}[f\psi^*]$. The total areas of blue and gray regions are equal, indicating energy balance.
(b) Block diagram illustrating the domain wall self-regulation process in the self-injection locking regime.
}
\end{figure}

To understand the system behavior, we first study the special case of zero dispersion (i.e., $\beta_2=0$). This simplifies the physical picture while not qualitatively impacting the results, and the inclusion of dispersion will be considered later. Zero dispersion removes the field derivative term from the LLE, which allows step discontinuities in the field. Such non-continuous-wave solutions will be shown to exist and are stable in the absence of dispersion. Since the continuous-wave operating point lies within the DI regime, fluctuations cause the field to destabilize away from the operating point. Fields in about half of the resonator will increase to the upper equilibrium while fields in the other parts decrease to the lower equilibrium. However, these local changes must still satisfy the laser locking condition. With respect to the pumping field, this occurs in a spatially averaged sense wherein the average intracavity field and power determines the operating point [Eq. (\ref{Eq_locking})]. The whole process is illustrated in Fig. \ref{FIG1}c. As an aside, the average field will change in response to the power changes, but such changes cannot flip the upper equilibrium to the lower equilibrium or vice versa, as such a spontaneous flipping of the field requires large fluctuations that are exponentially unlikely. 

In summary, beginning from the unstable branch continuous-wave operating point, the waveform evolves to a square-wave-like form that consists of sections of upper and lower equilibria. We will refer to these sections as high-field and low-field domains, respectively (Fig. \ref{FIG1}c). Between these domains, a field discontinuity occurs. Such discontinuities are known as domain walls (Fig. \ref{FIG1}c), analogous to the domain walls that separate magnetic domains in ferromagnetic materials. A similar optical concept, known as switching waves, has been extensively studied in fiber loops and resonators \cite{rozanov1982transverse, coen1999convection, parra2016origin}, and other names have been used as well, but the name ``domain wall'' is used here as we find it more convenient to describe the system with the spatial coordinate $\theta$ rather than using the ``fast time'' notation. It links to its topological origin as an object that continuously connects the high- and low-field domains, which possesses an exchange symmetry (see Appendix A) similar to the previous optical objects named domain walls \cite{haelterman1994polarization, zhang2009observation, zhang2010vector}. We avoid using the terminology ``dark soliton'' to describe the resulting waveform in the current system, as one may argue that the square-like wave does not occupy a localized region within the resonator, unlike the domain walls. For the special dispersionless case initially studied here, the domain walls have zero width due to absence of the derivative terms in the LLE. The width becomes finite for the normal GVD regimes as discussed below.

\section{Nonlinear injection locking with normal dispersion}

For the normal dispersion case where $\beta_2>0$, the domain formation process is qualitatively similar to the dispersionless case. The system still reaches the continuous-wave operating point followed by the emergence of high- and low-field domains. However, the walls at the boundary of the domains now have finite widths due to the dispersion term $\partial_\theta^2 \psi$ which imposes a continuity condition on the field. The spatial width of the domain wall is assumed to be much shorter than $2\pi$ (cavity round trip) such that boundary effects can be ignored. This will be discussed later in terms of domain wall interactions.

Typical domain wall solutions to the LLE (normal dispersion) are plotted in Fig. \ref{FIG1}d. The domain wall can be roughly divided into two parts. The portion close to the high-field domain has the form of a constant term minus an exponential that increases to the upper equilibrium, while the portion close to the low-field domain is either exponentially or oscillatory decaying to the lower field equilibrium. These behaviors are controlled by the eigenvalues of the field equation at the corresponding equilibria. At the upper and lower equilibria, the energy gain of the field equals the energy loss. For the upper (lower) part of the domain wall, the optical gain (cavity loss) term is more prominent, and the field has the tendency to converge to $\rho_\mathrm{H}$ ($\rho_\mathrm{L}$), expanding the high-field (low-field) domain. A stationary domain wall thus requires that these two effects balance each other. Quantitatively,
\begin{equation}
\int\left(-2|\psi|^2+2\mathrm{Re}[f\psi^*]\right)d\theta=0
\end{equation}
where the first term represents loss to the environment and the second term represents gain from the pump. If these tendencies are unbalanced, the domain wall will move as a whole in the direction determined by the dominant tendency. The overall speed of the domain wall can be calculated from the rate of energy change, and reads
\begin{equation}
D_1=\mp \frac{1}{|\rho_\mathrm{H}|^2-|\rho_\mathrm{L}|^2}\int_{-\infty}^\infty\left(-2|\psi|^2+2\mathrm{Re}[f\psi^*]\right)d\theta
\end{equation}
where the $\mp$ sign depends on which way the domain wall is oriented. Assuming the wave propagates to the right in the lab frame, the minus sign is taken if the high-field domain is also on the right of the domain wall, and vice versa. This expression can be interpreted as follows: after the domain wall moves in a unit time, the net effect is to pump the field in a unit length from low-field domain to high-field domain, and the energy difference is provided by the overall power absorbed by the domain wall. Despite the infinite integration limits, the integral converges due to the equilibrium state maintaining energy balance by itself, and the integrand converges to $0$ exponentially (or oscillating exponentially) at both sides.

With strong pumping, the domain wall converts pump energy to expand the high-field domain, while for weak pumping, loss causes the high-field domain to shrink. This dynamic process is also illustrated in Supplementary movies. For steady state operations, a critical $f$ value exists for a fixed detuning where the domain wall is in energy balance between pumping and loss. This value is known as the Maxwell point (MP) \cite{parra2016origin}, denoted as $f_\mathrm{MP}$, and plotted in Fig. \ref{FIG2}a. It can be determined by various analytical or variational methods. Near the critical point $\alpha=\sqrt{3}$, above which multiple equilibria can be found in the resonator, the MP can be obtained by asymptotic expansion (see Appendix A),
\begin{align}
f^2 &=\frac{8}{3\sqrt{3}}\left[1+\frac{\sqrt{3}}{2}\left(\alpha-\sqrt{3}\right)-\frac{3}{20}\left(\alpha-\sqrt{3}\right)^2\right.\nonumber\\
&+\left.\frac{999\sqrt{3}}{3500}\left(\alpha-\sqrt{3}\right)^3+O\left(\left(\alpha-\sqrt{3}\right)^4\right)\right],
\end{align}
which also defines a formal exchange symmetry of the domains and domain walls (see Appendix A). For intermediate $\alpha$ values, the MP can be estimated using variational methods based on the energy balance condition derived above (see Appendix B):
\begin{equation}
\label{Eq_MPVar}
f_\mathrm{MP}\approx\frac{4}{\pi^2}\alpha
\end{equation}

\section{Self-regulation of domain walls}

Normally, it is challenging to tune a pumping laser exactly to the MP so as to stop the domain wall from moving. However, because the self-injection locking process relates $\alpha$ to the intracavity field, it provides a feedback loop necessary to maintain laser lock to the MP. For example, suppose that the intracavity field has split into single low-field and high-field domains under constant pumping. Therefore, two oppositely-oriented domain walls appear in the system. If the pumping field is stronger than $f_\mathrm{MP}$ at the initial detuning, the expansion of the high-field domain will increase both the average field norm and average power in the resonator, which, in turn, increases the detuning according to the nonlinear locking relation. This brings the detuning closer to the MP, and the movement of domain walls slow down. Eventually the detuning converges to the MP, and the domain walls stop moving where the combination of average field and power maintain the appropriate detuning. The opposite situation of an initial pump field that is too low works in a similar way. To quantify the proportion of the high-field domain, we introduce the duty cycle variable, $w\%$, defined as the portion of the resonator with intracavity power higher than $|\rho_\mathrm{M}|^2$, which is analogous to the duty cycle describing square waves. The regulation process is summarized in Fig. \ref{FIG2}b. If other system parameters change, such as feedback phase and pump power, the above process is also capable of pulling the system to the Maxwell point defined by the new pump power.

As an aside, for the multiple-pulse case, the width for each individual dark pulse cannot vary independently but instead increase together with increasing pump power or vice versa. This behavior is determined by the movement of individual domain walls. For example, if pump power increases, then all domain walls with high-field domain to the left will shift right while all those with the opposite orientation will shift left, leading to an increase in all high-field domain lengths. It is therefore not possible to have complementary domain width changes, as this would require inconsistent domain wall movements for the same external pumping. As a result, the overall duty cycle of multiple pulses still follows the regulation process outlined in Fig. \ref{FIG2}b.

Domain walls are always generated in pairs with alternating orientations within a resonator subject to periodic boundary conditions. If a pair of domain walls is close enough, their exponential tails will overlap, leading to interactions between the domain walls. For the normal GVD case, domain walls attract each other when the high-field portions overlap, as the overlap integral leads to extra energy loss from the system (see Appendix C). This leads to collision and annihilation of the walls, and indicates that a bright-like pulse with $w\%$ close to $0\%$ is unstable in a normal GVD system. We point out that the presence of extra energy input channels can stabilize such bright-like pulses, known as ``platicons'' (e.g., the pump mode eigenfrequency can be red-shifted compared to the parabolic dispersion \cite{lobanov2015frequency} such that pumping becomes more efficient). For domain walls with overlapping oscillatory tails near the low-field domain, the interaction will be alternating between attraction and repulsion depending on the relative position of the tails. This results in multiple equilibrium positions of the two walls, and has been studied previously using bifurcation theory \cite{parra2016origin}. In the case of pumping the resonator with a fixed-detuning laser, if the pumping power is higher than $f_\mathrm{MP}^2$, the two domain walls will move towards each other until their low-field portions overlap, at which point they start to interact and settle into equilibrium, forming a localized structure referred to as ``dark soliton'' \cite{liang2014generation,xue2015mode}. If the pumping is too high, the maximum repulsion is not capable of holding the domain walls apart, leading to pair annihilation of the walls. Since the interaction between domain walls is limited before annihilation, dark solitons exist only within a very narrow region in the detuning-pump phase space \cite{godey2014stability}. From this point of view, the conventional dark solitons and platicons require the wall interactions to exist, which makes their duty cycles asymptotically close to $100\%$ and $0\%$, respectively. These interactions unify the domain wall picture with conventional, dissipative dark solitons as well as platicons generated with a fixed-detuning laser. On the other hand, domain walls in the nonlinear self-injection-locked resonator can be free from pairwise interactions, since the detuning is instead determined by the duty cycle and locked to the MP. As a result, the duty cycle can reach an intermediate value close to $50\%$.

\section{Feedback phase and comb efficiency}

\begin{figure*}
\includegraphics[width=170mm]{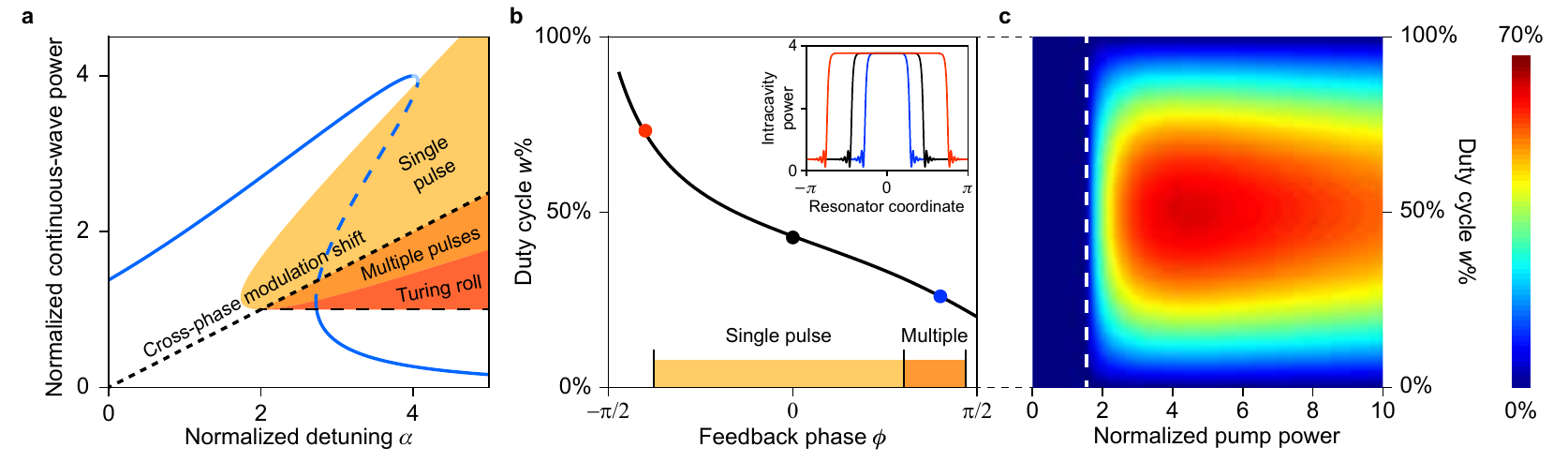}
\caption{\label{FIG3}
{\bf Control of pulse duty cycle and efficiency using feedback phase.}
(a) Phase diagram for the continuous-wave operating point with respect to normalized detuning $\alpha$ and normalized pump power $f^2$, showing regions corresponding to different pulse generation regimes. Single- or multiple-pulse generation processes will be preferred when the continuous-wave operating point is located in the corresponding colored region. The resonator pumping curve for $|f|^2=4$ (blue) is shown for comparison. Parameters within the MI region but outside the DI region leads to Turing rolls, where dark pulses can be generated but pulse number no longer depends on the mode number of the mode having the largest parametric gain.
(b) Duty cycle of the generated dark pulse as a function of feedback phase for pump strength $|f|^2=4$, assuming the domain wall width is negligible compared to the resonator circumference. Smaller feedback phase relative to $\phi=0$ leads to larger duty cycles, and vice versa. The range of feedback phase for this specific pump power that initiates single- or multiple-pulse generation is indicated at the bottom of the plot. The colored dots correspond to numerical data shown in the inset, and shows reasonable agreement with theoretical calculations. Inset: Simulated pulse profiles for different feedback phases, with $\phi=-0.4\pi$ (red), $\phi=0$ (black) and $\phi=0.4\pi$ (blue), at pumping strength $|f|^2=4$. Duty cycles obtained from the waveforms are 73\%, 43\%, and 26\% respectively and shown as dots in the main figure.
(c) Dependence of output comb efficiency (false color) on normalized pump power and duty cycle, assuming an overcoupling condition ($\kappa_\mathrm{ex}=(4/5)\kappa$). White dashed line indicates $|f|^2=8/(3\sqrt{3})$, the lower boundary for generating dark pulses.}
\end{figure*}

We now investigate the effects of the feedback phase, which will be shown to influence the number of dark pulses as well as the combined duty cycle of dark pulses. As shown previously, dark pulses form from fluctuations on the unstable branch. The growth of such fluctuations can be described by amplification of sidebands using the framework of MI. For a specific pair of sidebands with mode number $\pm m$ (relative to the pump mode), the MI region (where the parametric gain exceeds the cavity loss) has the same shape as that of the DI region for the pump mode in the $\alpha-\rho^2$ phase space, except that the detuning $\alpha$ is effectively red-shifted by $\beta_2 m^2/2$ (see Appendix D).
As the mode number of the emergent sideband pair determines the number of intensity peaks within the resonator, the number of domain wall pairs generated in the resonator will be close to the mode number of the sideband pair having the largest MI gain. This in turn depends on the $\alpha$ and $\rho$ coordinates of the continuous-wave operating point. Therefore the pulse number can be estimated given the continuous-wave operating point parameters (see Appendix D). We note that the exact pulse number is subject to domain wall collisions and other transient processes, and still has a certain degree of randomness.

In some cases, single pulse operation is desirable due to its smooth spectrum and the lack of uncertainty of the distance between different pulses. This requires the MI gain to monotonically decrease with $m$, such that the $m=1$ pair of modes experience the largest gain. This happens when the continuous-wave operating point is blue-detuned compared to the cross-phase modulation line $|\rho|^2=\alpha/2$ (see Appendix D). As such, the regions for the continuous-wave operating point favoring direct single- and multiple- pulse generation can be plotted (Fig. \ref{FIG3}a). We note that the continuous-wave operating point is implicitly dependent on the combinations of pumping strength and feedback phase. For each specific pumping strength, different segments of the pumping curve within each region can be converted to a specific range of the feedback phase (illustrated in Fig. \ref{FIG3}b). It is worth noting that dark pulses may also emerge after Turing rolls have formed through MI inside the resonator if the continuous-wave operating point is red-detuned compared to the DI region. Here the pulse number no longer depends on the mode number of the sideband pair with the largest MI gain, but on fluctuations of the Turing roll pattern envelope. A comparison of dark pulse generation in different regimes is given in Appendix E. 

Although the final detuning after pulse formation will be locked at the MP, the difference between the MP and the initial continuous-wave detuning will determine the duty cycle that is needed to adapt to this difference. If the width of the domain wall is negligible compared to the scale of the resonator, $\rho$ and $P$ can be approximated as the weighted average of the high- and low-field domain contributions. The duty cycle can thus be related to the feedback phase via the locking condition (see Appendix F) and, in principle, be solved numerically (Fig. \ref{FIG3}b). The duty cycle curve extends beyond the single pulse generation region, as the feedback phase can be tuned after the initial pulse formation to access single pulse operation for smaller feedback phase. Similarly, single pulse states can be achieved via phase tuning even if the direct formation process prefers multiple pulses at larger feedback phase. Increasing the pump power has the effect of extending the feedback phase ranges for initiating both single- and multiple-pulse generation. We note that $w\%$ becomes independent of the pulse number within the approximation of thin domain walls.

A practical application of controlling the duty cycle is to optimizing the overall comb power efficiency. Neglecting the domain wall widths, the output comb efficiency can be computed as
\begin{equation}
\frac{P_\mathrm{comb}}{P_\mathrm{in}}=w\%(1-w\%)\frac{|\rho_\mathrm{H}-\rho_\mathrm{L}|^2}{|f|^2}\frac{4\kappa_\mathrm{ex}^2}{\kappa^2}
\end{equation}
which is maximized at $w\%=50\%$ at fixed $|f|$ (Fig. \ref{FIG3}c). Efficiencies calculated using the waveforms from Fig. \ref{FIG3}b inset differ from the analytical results by less than 1\%. We note that for sufficiently small $\beta_2$, the comb power is mainly contributed by the domains and their associated power swings, therefore making it reasonable to ignore the domain wall contributions to the comb power.

\section{Imaging of domains}

\begin{figure}
\includegraphics[width=85mm]{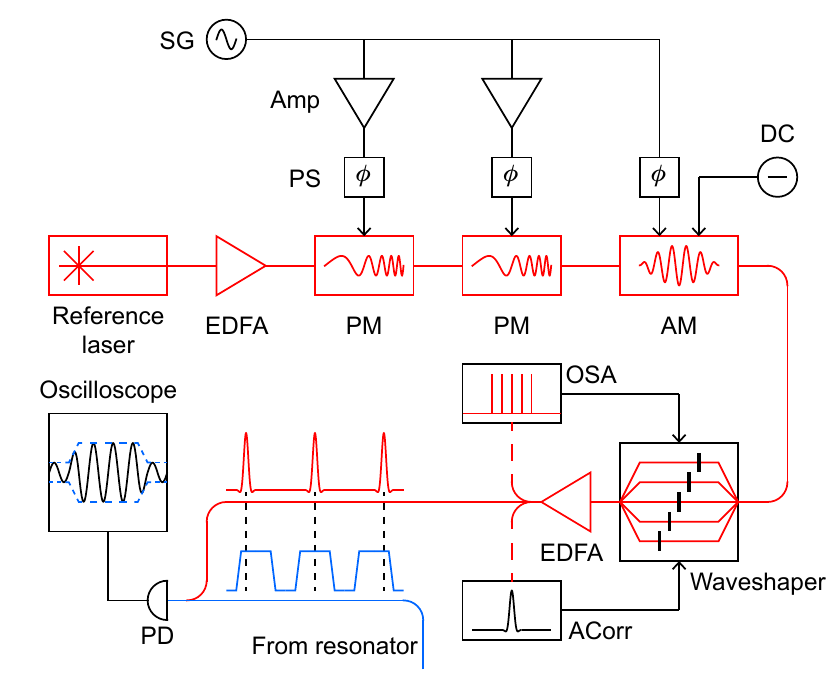}
\caption{\label{FIG4} {\bf Schematic of the electro-optic sampling measurement system.}
SG: analog radio-frequency signal generator. DC: direct-current voltage source. Amp: electrical amplifier. PS: electrical phase shifter. EDFA: erbium-doped fiber amplifier. PM: phase modulator. AM: amplitude modulator. OSA: optical spectrum analyzer. ACorr: auto-correlator. PD: photodetector. See \cite{yi2018imaging} for additional details.
}
\end{figure}

\begin{figure*}
\includegraphics[width=170mm]{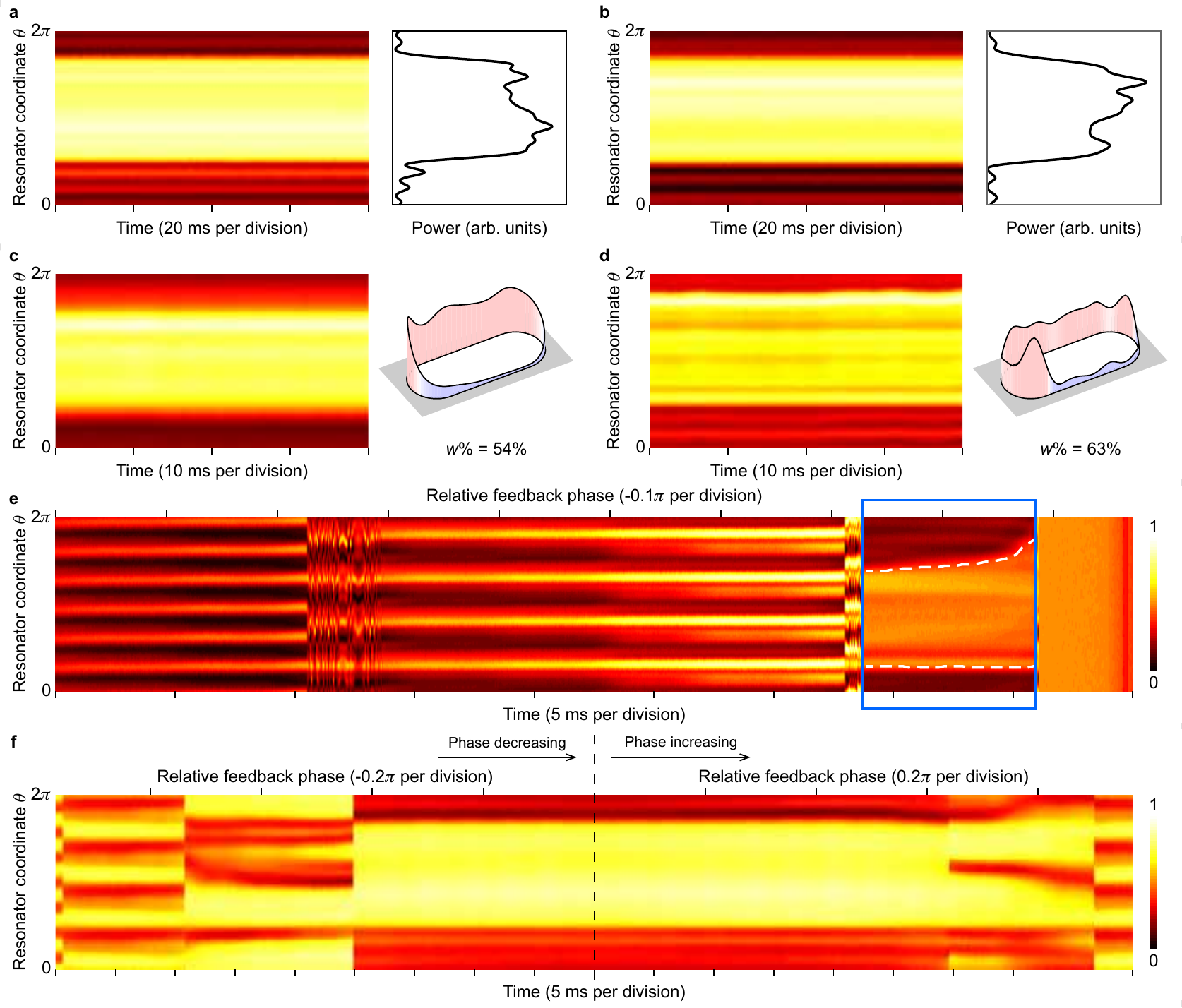}
\caption{\label{FIG5}
{\bf Measured domain wall images.}
(a) Left: time evolution plot for the intracavity field amplitude (false color) as a function of time with fixed feedback phase. The data are obtained by electro-optical sampling (Fig. \ref{FIG4}). In the plot, the vertical axis represents the angular coordinate for one round trip (0 to 2$\pi$) around the resonator. The horizontal axis is the evolution time, and each pixel column corresponding to one resonator round trip time. Slow drifts of the pulse have been removed for clarity. Right: averaged power profile for the measured waveform.
(b) Same as (a) but pumping a different longitudinal mode.
(c) Left: time evolution plot for the intracavity field amplitude (false color) as a function of time with fixed feedback phase. Another longitudinal mode is pumped compared to (a) and (b). Right: 3D representation of the averaged power profile on a racetrack resonator. The estimated duty cycle is also indicated below. 
(d) Same as panel (c), using the same longitudinal mode for pumping but with a slightly smaller feedback phase.
(e) Evolution plot for the intracavity field amplitude (false color) as a function of time and resonator coordinate measured while the feedback phase is decreasing. The estimated relative phase is derived from the applied piezoelectric voltage ($1.1\pi$ per volt). The single pulse section has been bounded by a blue box, within which the rising and falling edges of the pulse have been marked with white dashed lines.
(f) Same as (e) but the phase is decreasing for the first half of the scan and then increasing for the second half. The asymmetry of the phase ranges occupied by the single pulse is apparent.
}
\end{figure*}

We use the electro-optic sampling technique \cite{yi2018imaging} to experimentally obtain images of the domains and to verify some of the above theoretical predictions. A commercial InGaAsP distributed-feedback (DFB) laser around 1556 nm is endfire coupled without optical isolation to an integrated silicon nitride/silica resonator (free spectral range 10.85 GHz with no mode splittings observed in the vicinity of the pump mode) \cite{jin2021hertz}. The field is collected from the drop port of the resonator with a fiber lens to avoid the pumping field showing up in the results. The optical waveguide facets and lens fiber port are aligned by fine tuning a micro-positioner. The laser stage is equipped with piezoelectric position controls for all three translation degrees of freedom. For measurements with varied feedback phase, the gap between laser and resonator chip is tuned by applying a triangular voltage signal to the piezoelectric controller of the laser stage. The transduction factor is measured as $0.42$ $\mu$m V$^{-1}$, equivalent to about $1.1\pi$ feedback phase change per volt at 1556 nm. Changing the gap also weakly affects the coupling efficiency between the laser and resonator, which is estimated to be $<0.5$ dB for the tuning range used. It is noted that implementation of a heater section on the waveguide can enable on-chip thermal control of feedback phase \cite{xiang2021laser}.

Pulse snapshot images are obtained by mixing at a photodetector the dark pulse train with an electro-optically (EO) generated comb having a slightly different repetition rate. The electro-optic sampling measurement is illustrated in Fig. \ref{FIG4}. An electro-optical sampling pulse stream is generated by two phase modulators and one amplitude modulator followed by amplification using an erbium-doped fiber amplifier (EDFA). The output is then conditioned by a waveshaper to form the sampling pulse stream. The corresponding electro-optic comb spectrum is measured by an optical spectrum analyzer (OSA). Individual comb line amplitudes are then adjusted using the waveshaper to tailor the comb spectrum. The resulting comb has around 40 lines with equal intensity (variation $<1$ dB). The comb is also characterized by an autocorrelator in the time domain, and the result is used to adjust the dispersion applied using the waveshaper. The dark pulse collected via the fiber lens is mixed with the electro-optic sampling pulses on a photodetector (1 GHz bandwidth). The radio frequency signal is then collected by the oscilloscope, digitally demodulated and segmented. The segmentation length is variable and determined from the waveform to maintain the periodicity of the pulse and to correct for repetition rate drifting. Each piece of waveform is then down-sampled to 128 points for plotting and averaging.

By pumping different longitudinal modes, pulse states can be observed in the resonator (Fig. \ref{FIG5}a and b). The square-like waveform is apparent from the time evolution plot and its 3D representation (Fig. \ref{FIG5}c and d). The variations of the field in both the high-field and low-field domains are believed to result from resonator inhomogeneity along the propagating direction (see Appendix G) as well as inaccuracies in the sampling process. As $\rho_\mathrm{M}$ cannot be accurately retrieved from the experiment, the pulse width here is determined instead as the portion with an optical power greater than the average of the 87.5\% and 12.5\% quantiles of the round-trip waveform. 
For different feedback phases, we are able to observe pulse states with different duty cycles (Fig. \ref{FIG5}c and d). In strong contrast to previously demonstrated bright solitons, dark solitons or platicons, the measured pulse width occupies a significant portion of the resonator.

We have also swept the feedback phase by adjusting the coupling gap between the laser and resonator, and monitored the evolving field in the resonator during the scanning process (Fig. \ref{FIG5}e). When the feedback phase is decreasing, Turing rolls, breathing states and dark pulse states can be observed during the single scan. Notably, the pulse width for a single pulse state near the end of the scan visibly widens (highlighted in Fig. \ref{FIG5}e). For the central region of Fig. \ref{FIG5}e which consists of four dark pulses, the increase of field intensities also indicates that the duty cycle is increasing. These observations are in qualitative agreement with Fig. \ref{FIG3}a and \ref{FIG3}b and consistent with MP predictions. Deviations of these measured results from the ideal domain wall shape are believed to be related to the distributed backscattering in the resonator (see Appendix G). As the phase decreases during the scan, the phase range for multiple pulses appears longer than that of single pulse. This happens because it is possible for multiple pulses, once formed, to exist in the single-pulse initiation range. Additional measurements including both phase scanning directions have been performed (Fig. \ref{FIG5}f), where the asymmetry of the states with respect to the scan direction indicates such hysteresis behavior of the pulses. We note that the duty cycle change of the single pulse is not obvious from the plot. This is believed to result from the large dispersion $\beta_2$ for the resonator used, which significantly increases the interaction between domain walls (see Appendix G).

\begin{figure}
\includegraphics[width=85mm]{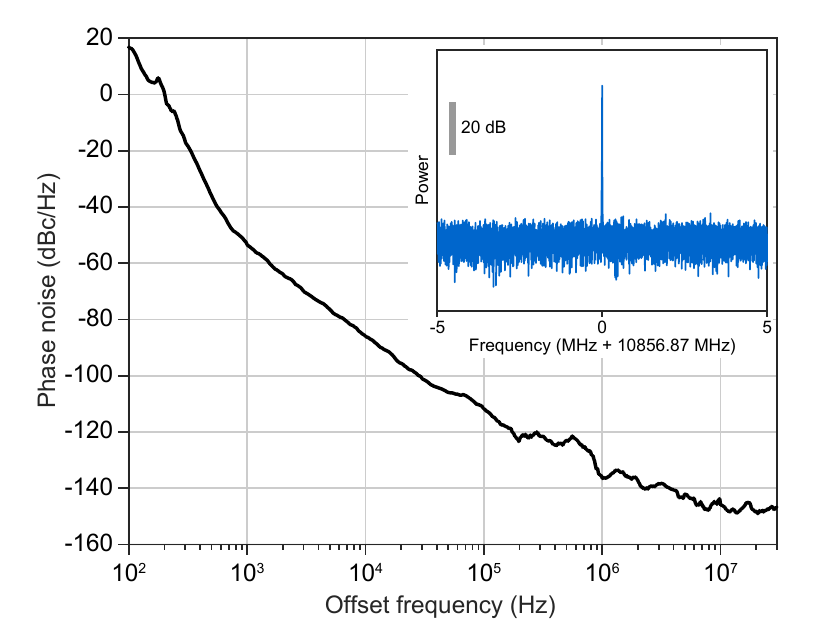}
\caption{\label{FIG6}
{\bf Representative microwave phase noise spectrum of the dark pulse repetition rate signal.}
Inset shows the measured repetition rate electrical spectrum with a resolution bandwidth of 1 kHz.
}
\end{figure}

Noise spectra of the pulse repetition rate signal have also been measured, and a representative noise spectrum is shown in Fig. \ref{FIG6}. At 10 MHz offset frequencies, typical phase noise observed are $-140$ to $-145$ dBc/Hz and are comparable with previous bright soliton systems with 10-GHz-scale repetition rates \cite{yang2021dispersive}. We note that a ``quiet'' operation point \cite{yi2017single}, where repetition rate noise is significantly compressed compared with normal operations, should also be possible in the present systems. 

\section{Summary and Discussion}

In summary, the formation dynamics of dark pulses in microresonators via the self-injection-locking process have been analyzed in terms of domain walls. The resulting system has a new physical property associated with self-regulation of the domain walls. The nonlinear waves were also imaged using an electro-optic sampling system, and the measurements verified predictions of the model. Self-regulation allows operation of the normal dispersion microcomb at previously difficult-to-access duty cycles that offer high power efficiency for comb states \cite{xue2017microresonator} as well as for microwave generation \cite{jin2021hertz}. The duty cycle is controlled by the feedback phase, which in future designs could be electrically varied using, for example, an on-chip heater \cite{xiang2021laser} or a phase control section added to the III-V laser.

There are many other effects that can be included in the model and these may lead to new phenomena in the system. For example, normalized backscattering of the resonator is heavily dependent on the geometry and fabrication details and may range from $10^{-3}$ to $10^1$. Strong backscattering causes mode splitting and, when combined with optical nonlinearity, can lead to new modal dynamics. There have also been numerical efforts to generalize the backscattering to each pair of longitudinal modes \cite{kondratiev2020modulational}. For Rayleigh scattering, the actual behavior of backscattering becomes more complex, where large amplitude and phase variations across different pairs of modes may be observed \cite{jin2021hertz}. For controllable coupling strength and frequency location of the split resonance, gratings can be introduced to the resonator structure \cite{yu2021spontaneous}. High-order dispersion can also be added to the model, and becomes important as the second-order dispersion approaches zero, Here, similar domain-wall-like behavior of the pulse has also been shown in the numerical simulations \cite{anderson2020zero}.

\section*{Acknowledgements}

The authors thank Q. Yang and L. Wu for discussions. Funding is provided by the Defense Advanced Research Projects Agency (DARPA) under A-PhI (FA9453-19-C-0029) and DODOS (HR0011-15-C-055) programs and the Air Force Office of Scientific Research (AFOSR) (FA9550-18-1-0353).


\appendix

\begin{widetext}

\section{Calculation of the Maxwell point (Asymptotic approach) and construction of the exchange symmetry}

Here we calculate the Maxwell point analytically by expressing the domain wall solution and the corresponding pump as an asymptotic series around the critical point $\alpha=\sqrt{3}$. The method is based on the multiple-scale analysis \cite{coen1999convection} previously applied to fiber systems. Starting from the stationary LLE,
\begin{equation}
0 = -(1+i\alpha)\psi-i\frac{\beta_2}{2}\frac{\partial^2 \psi}{\partial\theta^2}+i|\psi|^2\psi+f
\end{equation}
we substitute $\alpha=\sqrt{3+\epsilon^2}$, where $\epsilon$ will be used as the formal expansion parameter. The appearance of $\epsilon^2$ in $\alpha$ takes account of the pitchfork bifurcation near the critical point and we can restrict the expansion to integer powers of $\epsilon$. The pump term can be expanded as
\begin{equation}
f=\frac{\sqrt{8}}{\sqrt[4]{27}}\left(1+\frac{1}{2!}\frac{\epsilon^2}{4}+\frac{1}{4!}f_4\epsilon^4+\frac{1}{6!}f_6\epsilon^6+O(\epsilon^8)\right)
\end{equation}
where only even-order terms have been retained as $f$ should be a single-valued function with respect to $\alpha$, and $f_2=1/4$ has been calculated directly as the Maxwell point line must be tangent with the DI boundaries on the phase diagram. The $f_4$ and $f_6$ terms will be calculated using the expansion of the domain wall solution:
\begin{align}
\psi & = u + iv\\
u & =\frac{3}{4}\frac{\sqrt{8}}{\sqrt[4]{27}}\left(1+u_1\epsilon+\frac{1}{2!}u_2\epsilon^2+\frac{1}{3!}u_3\epsilon^3+\frac{1}{4!}u_4\epsilon^4\right)+O\left(\epsilon^5\right)\\
v & =-\frac{\sqrt{3}}{4}\frac{\sqrt{8}}{\sqrt[4]{27}}\left(1+v_1\epsilon+\frac{1}{2!}v_2\epsilon^2+\frac{1}{3!}v_3\epsilon^3+\frac{1}{4!}v_4\epsilon^4\right)+O\left(\epsilon^5\right)
\end{align}
where $u$ and $v$ are the real and imaginary parts of $\psi$, and $u_j$ and $v_j$ ($j=1,2,3,4$) are real functions that represent the expanded field at various orders. We also define a scaled position parameter:
\begin{equation}
x=\frac{\left|\epsilon\right|\theta}{\sqrt[4]{3}\sqrt{2\beta_2}}
\end{equation}
The scaling contains $\epsilon$ and takes account of the expansion of domain wall width near the critical point. Using the scaled position, we separate the real and imaginary part of the equations:
\begin{align}
f-u-v(u^2+v^2-\alpha) +\frac{\epsilon^2}{4\sqrt{3}}\frac{d^2v}{dx^2} & =0\\
-v+u(u^2+v^2-\alpha)  -\frac{\epsilon^2}{4\sqrt{3}}\frac{d^2u}{dx^2} & =0
\end{align}
The structure of the equation pair leads to a staggered expansion scheme. To determine $u_j$ and $v_j$, the imaginary part of LLE needs to be expanded to $\epsilon^j$, while the real part should be expanded to $\epsilon^{j+2}$. Calculating $u_j$ and $v_j$ also leads to the value of the $f_{j+2}$ coefficients.

At $\epsilon^1$ order of the imaginary part, we obtain
\begin{equation}
u_1+v_1=0\ \ \rightarrow\ \ u_1=-v_1
\end{equation}
Substituting into the $\epsilon^3$ order of the real part, we obtain
\begin{equation}
-2v_1+8v_1^3-v_1''=0
\end{equation}
where prime denotes derivative with respect to $x$. This equation resembles the LLE but with loss and pump terms removed. Its fundamental dark solution can be found as
\begin{equation}
v_1=\frac{1}{2}\mathrm{tanh}\,x\equiv\frac{\eta}{2}
\end{equation}
where we introduced the shorthand notation $\eta\equiv\mathrm{tanh}\,x$.

We proceed to the $\epsilon^2$ order of the imaginary part:
\begin{equation}
-1+2\eta^2+2u_2+2v_2=0\ \ \rightarrow\ \ u_2=\frac{1}{2}\left(1-2\eta^2-2v_2\right)
\end{equation}
Substituting into the $\epsilon^4$ order of the real part, we obtain
\begin{equation}
4f_4+\frac{13}{4}-\frac{27}{2}\eta^2+18\eta^4+6(3\eta^2-1)v_2-3v_2''=0
\end{equation}
This is a Legendre differential equation in $\eta$ after substituting $d/dx\rightarrow(1-\eta^2)d/d\eta$, and its general solution is the associated Legendre polynomial $P_2^2(\eta)=3(1-\eta^2)=3\sech^2x$. The appearance of this term with undetermined coefficients is not surprising as the domain wall has translational invariance, and adding the term $\sech^2x=(\tanh x)'$ simply shifts the domain wall up to $\varepsilon^2$ order. Here we will choose $v_2(x=0)=u_2(x=0)=1/4$ to fix the coefficient. The $f_4$ appears as an eigenvalue of the differential equation that prevents the special solution to be divergent as $\eta\rightarrow\pm1$ (equivalently $x\rightarrow\pm\infty$). With these considerations, the special solution can be solved as
\begin{equation}
v_2=-\frac{9}{20}+\frac{7}{10}(1-\eta^2)-\frac{3}{5}(1-\eta^2)\ln(1-\eta^2)
\end{equation}
and we find that $f_4=-47/80$.

We summarize the rest of the expansion results below without detailed calculation procedures:
\begin{equation}
u_3=\frac{1}{10}\left(-15\eta+18\eta^3-36\eta(1-\eta^2)\ln(1-\eta^2)-10v_3\right)
\end{equation}
\begin{equation}
v_3=\frac{151}{200}\eta+\frac{111}{50}\eta(1-\eta^2)-\frac{77}{50}(1-\eta^2)x-\frac{9}{25}\eta(1-\eta^2)\ln(1-\eta^2)+\frac{27}{25}\eta(1-\eta^2)\ln^2(1-\eta^2)
\end{equation}
\begin{align}
u_4=\frac{1}{200}&\left(
-715+10220\eta^2-9048\eta^4-2464\eta(1-\eta^2)x-720(1-\eta^2)\ln(1-\eta^2)+6048\eta^2(1-\eta)^2\ln(1-\eta^2)\right. \nonumber\\
& \left.-864(1-\eta^2)\ln^2(1-\eta^2)+2592\eta^2(1-\eta^2)\ln^2(1-\eta^2)-200w_4\right)
\end{align}
\begin{align}
w_4 & =\frac{18027}{14000}-\frac{46133}{7000}(1-\eta^2)+\frac{6204}{875}(1-\eta^2)^2-\frac{16008}{875}(1-\eta^2)\ln(1-\eta^2)+\frac{1782}{125}(1-\eta^2)^2\ln(1-\eta^2) \nonumber\\
& -\frac{108}{25}(1-\eta^2)\ln^2(1-\eta^2)+\frac{486}{125}(1-\eta^2)^2\ln^2(1-\eta^2)-\frac{216}{125}(1-\eta^2)\ln^3(1-\eta^2)+\frac{324}{125}(1-\eta^2)^2\ln^3(1-\eta^2)\nonumber\\
& -\frac{154}{125}(1-\eta^2)\eta x+\frac{924}{125}(1-\eta^2)\eta x\ln(1-\eta^2)
\end{align}
\begin{equation}
f_6=\frac{95027}{11200}
\end{equation}
These results can be verified with the help of computer algebra systems. Collecting the $f_j$ coefficients and expressing them using $\alpha$ leads to our final result:
\begin{equation}
f^2=\frac{8}{3\sqrt{3}}\left[1+\frac{\sqrt{3}}{2}\left(\alpha-\sqrt{3}\right)-\frac{3}{20}\left(\alpha-\sqrt{3}\right)^2+\frac{999\sqrt{3}}{3500}\left(\alpha-\sqrt{3}\right)^3+O\left(\left(\alpha-\sqrt{3}\right)^4\right)\right]
\end{equation}

Although the procedure can be used to calculate arbitrarily high-order terms, its usefulness for calculation the Maxwell point away from the critical point is limited. Just above $\alpha=2$ the second-order term becomes smaller than the third-order term, indicating a truncation error of about $1\%$, and larger detunings further increase the error. It is not known if the above series has a finite radius of convergence.

We observe that all fields at odd orders of $\epsilon$ is odd in $x$ (and therefore $\theta$), while all fields at even orders of $\epsilon$ is even in $x$. This indicates that the solution respects the symmetry of the equation, and remains invariant under $(\epsilon,\theta)\rightarrow (-\epsilon,-\theta)$. The $\theta\rightarrow -\theta$ operation alone flips the orientation of the domain wall, therefore the $\epsilon\rightarrow -\epsilon$ establishes a formal exchange symmetry of the domain wall solution. Specifically, the high-field domain ($\rho_\mathrm{H}$) is mapped to the low field domain ($\rho_\mathrm{L}$) and vice versa through $\epsilon\rightarrow -\epsilon$. Corresponding points on the domain wall interior can also be mapped to each other. Although domain symmetries can be realized by other means (e.g. through the permutation group on the roots of pumping curve polynomial), the expansion parameter $\epsilon$ provides a way to continuously connect the different domain wall states and serves as an order parameter of the system.

The symmetry argument can be generalized to the cases when the pump is away from the Maxwell point and the domain wall is moving. We add a speed term $D_1$ to the LLE, which now reads,
\begin{equation}
-D_1 \frac{\partial \psi}{\partial \theta} = -(1+i\alpha)\psi-i\frac{\beta_2}{2}\frac{\partial^2 \psi}{\partial\theta^2}+i|\psi|^2\psi+f
\end{equation}
The rescaling of $D_1$ reads
\begin{equation}
d_1 = \frac{D_1}{\sqrt[4]{3} \sqrt{2 \beta_2} \left|\epsilon\right|}
\end{equation}
Now the pump expansion may contain odd orders of $\epsilon$:
\begin{equation}
f=\frac{\sqrt{8}}{\sqrt[4]{27}}\left(1+\frac{1}{2!}\frac{\epsilon^2}{4}+\frac{1}{3!}f_3\epsilon^3+\frac{1}{4!}f_4\epsilon^4+O(\epsilon^5)\right)
\end{equation}
where fewer terms have been taken due to the complexity of the expressions. Following the same procedures, we find that
\begin{equation}
u_1=d_1 - \frac{1}{2}\sqrt{1-12{d_1}^2} \eta
\end{equation}
\begin{equation}
v_1=-d_1 + \frac{1}{2}\sqrt{1-12{d_1}^2} \eta
\end{equation}
\begin{equation}
f_3=\frac{3}{2}d_1 - 24 {d_1}^3
\end{equation}
\begin{equation}
\begin{aligned}
u_2&=-\frac{1}{20}+\frac{24}{5}{d_1}^2 - \frac{4d_1(1-36{d_1}^2)}{5\sqrt{1-12{d_1}^2}}\eta + \left(\frac{3}{10}-\frac{34}{5} d_1^2\right)(1-\eta^2)\\ &+\frac{3}{5}  (1-12{d_1}^2)(1-\eta^2) \ln{(1-\eta^2)} -\frac{12}{5}d_1(1-20{d_1}^2)(1-\eta^2)x
\end{aligned}
\end{equation}
\begin{equation}
\begin{aligned}
v_2&=-\frac{9}{20} + \frac{16}{5} {d_1}^2 + \frac{24d_1(1-16{d_1}^2)}{5 \sqrt{1-12{d_1}^2}}\eta + \left(\frac{7}{10}-\frac{26}{5} {d_1}^2\right) (1-\eta^2) \\ &-\frac{3}{5}  (1-12{d_1}^2)(1-\eta^2) \ln{(1-\eta^2)} +\frac{12}{5}d_1(1-20{d_1}^2)(1-\eta^2)x
\end{aligned}
\end{equation}
\begin{equation}
f_4=\frac{1}{80}\left(-47+2688 {d_1}^2 - 55296{d_1}^4\right)
\end{equation}
where $\eta$ is now redefined as $\eta \equiv \tanh{\left(\sqrt{1-12d_1^2} x\right)}$. The solution is invariant under $(\epsilon,\theta,D_1) \rightarrow (-\epsilon,-\theta,-D_1)$. As flipping orientation of the domain wall changes the sign of $\theta$ and $D_1$, the $\epsilon$ parameter again connects the different domain wall states continuously.

\section{Calculation of the Maxwell point (Variational approach)}

\begin{figure*}
\includegraphics[width=170mm]{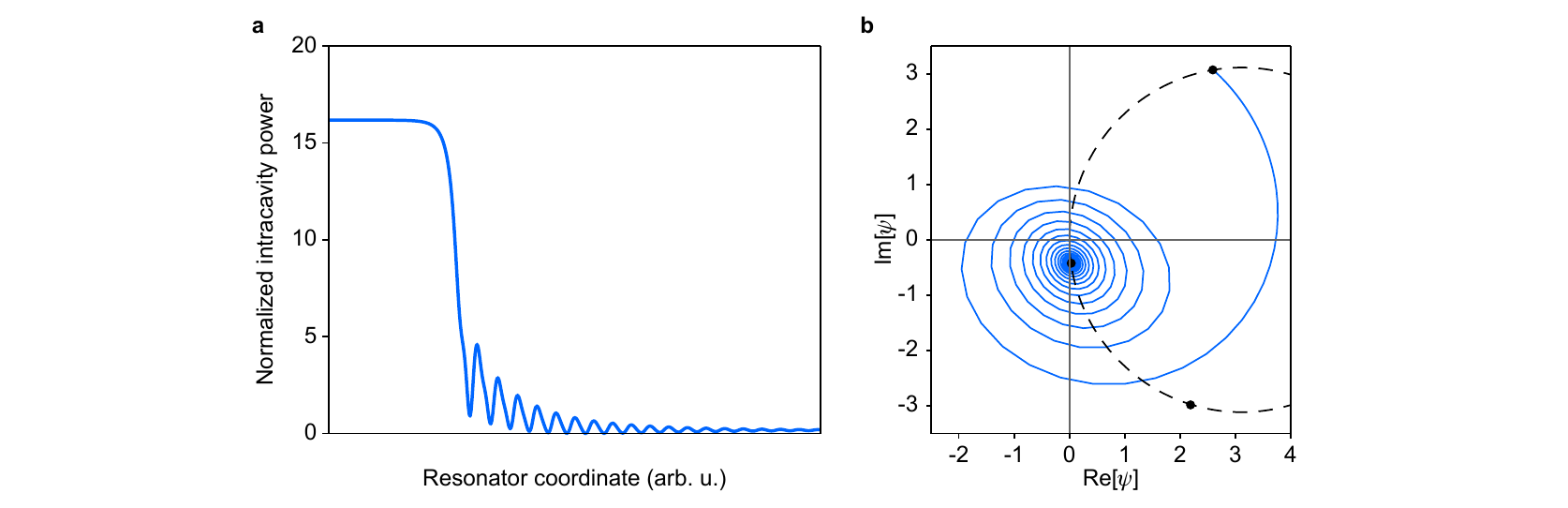}
\caption{\label{FIG7}
Domain wall solution for $\alpha=15$.
(a) The solution represented in the spatial domain.
(b) The solution represented in the complex $\psi$ plane, connecting $\rho_\mathrm{L}$ to $\rho_\mathrm{H}$ (blue). The three black dots mark the equilibrium values, $\rho_\mathrm{H}$ (top), $\rho_\mathrm{M}$ (bottom), and $\rho_\mathrm{L}$ (left). They are all on the energy balance circle (black dashed curve), described by $-2|\psi|^2+2\mathrm{Re}[f\psi^*]=0$. Within the circle the field experiences net gain, and outside the circle the field experiences net loss.
}
\end{figure*}

Here we estimate the Maxwell point for intermediate detuning levels based on the spatial characteristics of the domain wall solutions. The domain wall solution for $\alpha=15$ is represented in Fig. \ref{FIG7}. In the complex $\psi$ plane, the low-field section coils around $\rho_\mathrm{L}$ and spirals outwards, while the high-field section converges towards $\rho_\mathrm{H}$ exponentially. The energy balance condition, $-2|\psi|^2+2\mathrm{Re}[f\psi^*]=0$, has the shape of a circle on the complex $\psi$ plot and indicates if the local field is gaining or losing energy. For the low-field section, the field alternates between net gain and net loss, but the average effect is loss; while for the high-field section, the entire field has net gain that balances the loss from low-field sections.

We first approximate the spiral section with an exponential function. We will start from the origin and ignore the nonlinear term. We will also drop the pump term as gain is not important here compared to loss. The LLE is now approximated as
\begin{equation}
0=-(1+i\alpha)\psi-i\frac{\beta_2}{2}\frac{\partial^2\psi}{\partial\theta^2}
\end{equation}
Solving the linearized equation, $\psi$ for the low-field section can be approximated as
\begin{equation}
\psi\approx c_0\exp\left(i\sqrt{\frac{2\alpha}{\beta_2}}\theta+\frac{\theta}{\sqrt{2\alpha\beta_2}}\right)
\end{equation}
where we approximated $\sqrt{1-i/\alpha}$ with $1-i/(2\alpha)$. We take $\theta=0$ to be the point where $|\psi|$ reaches the middle equilibrium $|\rho_\mathrm{M}|$. $\rho_\mathrm{M}$ is approximated to be $\sqrt{\alpha}$, which is the power required for the Kerr effect to compensate for the detuning, and we have $|c_0|\approx\sqrt{\alpha}$. Now the loss on the low-field section can be calculated as
\begin{equation}
\int 2|\psi|^2 d\theta \approx 2 \int_{-\infty}^{0} \alpha \exp\left(\frac{2\theta}{\sqrt{2\alpha\beta_2}}\right)d\theta = 2 \frac{\sqrt{2\alpha\beta_2}}{2} \alpha=\sqrt{2\beta_2}\alpha^{3/2}
\end{equation}

For the connecting part, we again approximate it as an exponential by including the nonlinear effects and ignore the loss. The overall effect is to replace $\sqrt{2\alpha/\beta_2}$ with $\sqrt{2(\alpha-\rho)/\beta_2}$. We approximate $\alpha-\rho\approx 1$ at the start of the connecting region, with
\begin{equation}
\psi\approx \sqrt{\alpha}\exp\left(i\sqrt{\frac{2}{\beta_2}}\theta\right)
\end{equation}
This part connects to the high-field section, where $\mathrm{Arg}[\psi]$ approaches $\mathrm{Arg}[\rho_\mathrm{H}]\approx\pi/2$ exponentially while $|\psi|$ is approximately constant. We therefore approximate $\psi$ as
\begin{equation}
\psi\approx \sqrt{\alpha}\exp\left[i\frac{\pi}{2}\left(1-\exp\left(-\frac{2}{\pi}\sqrt{\frac{2}{\beta_2}}\theta\right)\right)\right]
\end{equation}
where the inner exponent is chosen to continuously match the connecting part. Although this exponent does not match the eigenvalue near $\rho_\mathrm{H}$, and the resulting asymptotic behavior is different, the energy gain is concentrated near the $|\psi|\approx |\rho_\mathrm{M}|$ section instead of the tails, and we estimate the overall gain using the approximated shape as
\begin{equation}
\int 2\mathrm{Re}[f\psi^*] d\theta \approx 2f\sqrt{\alpha} \int \sin\left[\frac{\pi}{2}\exp\left(-\frac{2}{\pi}\sqrt{\frac{2}{\beta_2}}\theta\right)\right] d\theta
\approx 2f\sqrt{\alpha}\times \frac{\pi}{2}\sqrt{\frac{\beta_2}{2}} \int_0^\infty \frac{\sin(\pi z/2)}{z} dz
=\frac{\pi^2}{4}f\sqrt{2\beta_2}\sqrt{\alpha}
\end{equation}
where the substitution of $z=\exp[-(2/\pi)\sqrt{2/\beta_2}\theta]$ is used and we have extended the integration limit to infinity. As the domain wall requires that gain equals loss so as to remain stationary, we can equate the gain and loss approximately:
\begin{equation}
\sqrt{2\beta_2}\alpha^{3/2}=\frac{\pi^2}{4}f\sqrt{2\beta_2}\sqrt{\alpha}
\end{equation}
The $\beta_2$ cancels out as expected, and we are left with
\begin{equation}
f\approx\frac{4}{\pi^2}\alpha
\end{equation}

Although the estimation used various approximations, the overall agreement to the numerically obtained result is rather satisfactory, achieving a minimum pump amplitude error of 2.4\% (pump power error $4.8\%$) at $\alpha\approx 16$, and maintaining amplitude error less than $10\%$ within the range of $10<\alpha<50$. We note that the Maxwell point is not well-defined for arbitrarily large $\alpha$, as the domain wall starts to breathe for the expected energy balance condition after around $\alpha>85$.

\section{Interactions of the domain wall}

Here we consider the domain wall interactions by studying the energy balance of two domain walls that are within proximity of each other. Assume first that a bright pulse is formed consisting of two domain walls with its high-field section facing the center and low-field section extending to infinity. If the domain wall solution is denoted as $\psi_\mathrm{DW}(\theta)$ (with its low-field section on the left), then the bright pulse can be approximated as
\begin{equation}
\psi=\rho_\mathrm{H}
+[\psi_\mathrm{DW}(\theta+\theta_\mathrm{DW})-\rho_\mathrm{H}]
+[\psi_\mathrm{DW}(-\theta+\theta_\mathrm{DW})-\rho_\mathrm{H}]
=\psi_\mathrm{DW}(\theta+\theta_\mathrm{DW})+\psi_\mathrm{DW}(-\theta+\theta_\mathrm{DW})-\rho_\mathrm{H}
\end{equation}
where the first (second) bracket describes the left (right) domain wall and $\theta_\mathrm{DW}$ describes the position of the domain wall.

Each term in the expanded expression of $\psi$ (two domain walls, one equilibrium background) can maintain its own energy balance when the other terms are absent. However, their co-existence leads to cross terms and breaks the energy balance:
\begin{equation}
\frac{\partial}{\partial\tau}\int_{-\infty}^{\infty} |\psi|^2 d\theta
=\int_{-\infty}^{\infty}
\left(-2|\psi|^2+2\mathrm{Re}[f\psi^*]\right)d\theta
=-4\int_{-\infty}^{\infty}\mathrm{Re}\{[\psi_\mathrm{DW}(\theta+\theta_\mathrm{DW})-\rho_\mathrm{H}]^*
[\psi_\mathrm{DW}(-\theta+\theta_\mathrm{DW})-\rho_\mathrm{H}]\}d\theta
\end{equation}

For domain walls that are separated by a sufficiently long distance, the main contribution of the integral comes from the overlapping high-field tails. Since the two domain walls share the same shape, the overlap integral is positive, indicating the composite system will lose energy and shrinks the high-field domain. This can also be interpreted as an attracting force between the two walls.

The analysis is similar for a dark pulse with the low-field section of the two domain walls facing the center, except that the low-field tail of the wall may become oscillatory. In this case the overlap integral may be positive or negative depending on the relative position of the tails. Accordingly, the domain wall interactions with overlapping low-field portions are either attractive or repulsive.

\section{Equivalence of DI and MI, and the number of dark pulses}

In the main text, the formation of domain walls has been described in the spatial domain using DI, i.e. fields on the unstable branch evolve to the higher or lower stable branches. However, the process of comb formation has been better understood in the frequency domain in terms of MI, where signal and idler sidebands experience net positive gain when the pump mode power is above a certain threshold \cite{chembo2010modal,godey2014stability}. We will first reconcile the DI and MI concepts, which will be helpful for constructing a geometrical representation of the effects from feedback phase, and then proceed to estimate the number of dark pulses.

For the zero dispersion case, the parametric gain for the continuous-wave state can be found from a standard perturbation analysis. Define $\delta\psi$ as the perturbation of the field. Linearizing around the equilibrium $\psi=\rho$, we arrive at the coupled equations for the perturbation:
\begin{equation}
\frac{\partial}{\partial\tau}
\begin{pmatrix}
\delta\psi \\ \delta\psi^*
\end{pmatrix}
=
\begin{pmatrix}
-(1+i\alpha-2i|\rho|^2) & i\rho^2 \\
-i(\rho^*)^2 & -(1-i\alpha+2i|\rho|^2) 
\end{pmatrix}
\begin{pmatrix}
\delta\psi \\ \delta\psi^*
\end{pmatrix}
\end{equation}
The parametric gain is then the larger eigenvalue of the coefficient matrix:
\begin{equation}
\lambda=-1+\sqrt{|\rho|^4-(\alpha-2|\rho|^2)^2}
\end{equation}
The gain becomes positive within the DI boundary and negative outside the boundary, consistent with the hysteresis theory. For a fixed $|\rho|^2$, the gain is the largest at $\alpha=2|\rho|^2$ (e.g. phase matching occurs when cross-phase modulation is compensated), and becomes smaller as the detuning moves away from this optimal value.

For the case with dispersion, we assume the perturbation is in the form $\delta\psi=\delta\psi_+e^{im\theta}+\delta\psi_-e^{-im\theta}$, where $\delta\psi_\pm$ are mode amplitudes and $m$ is the undetermined relative mode number for the perturbation. As the exponential functions are eigenfunctions in systems with translational symmetry, this form of perturbation ensures that the small-signal gain can be well defined. Linearizing around $\psi=\rho$ and separating the $e^{\pm im\theta}$ components, we get:
\begin{equation}
\frac{\partial}{\partial\tau}
\begin{pmatrix}
\delta\psi_+ \\ \delta\psi_-^*
\end{pmatrix}
=
\begin{pmatrix}
-(1+i\alpha-i\zeta-2i|\rho|^2) & i\rho^2 \\
-i(\rho^*)^2 & -(1-i\alpha+i\zeta+2i|\rho|^2) 
\end{pmatrix}
\begin{pmatrix}
\delta\psi \\ \delta\psi_-^*
\end{pmatrix}
\end{equation}
where $\zeta=\beta_2m^2/2$ is the four-wave-mixing phase mismatch. Comparison with the DI calculations shows that perturbation on the signal-idler waves is formally equivalent to perturbations on the pump mode itself ($\delta\psi_+ \leftrightarrow \delta\psi$ and $\delta\psi_-^* \leftrightarrow \delta\psi^*$), but with the detuning shifted by $\zeta$. As a result, the parametric gain in this case is modified as
\begin{equation}
\lambda=-1+\sqrt{|\rho|^4-(\alpha-\zeta-2|\rho|^2)^2}
\end{equation}
reproducing the previous results \cite{godey2014stability}.

For a given $m$ number, the instability criteria can be geometrically represented by shifting the original DI region by $\zeta$ horizontally on the $\alpha-\rho^2$ plot. If the dispersion is anomalous ($\zeta<0$), the region would sweep to the blue side, including the upper branches with $|\rho|>1$, recovering the conventional MI results. Here we are interested in normal dispersion ($\zeta>0$), where the region sweeps to the red side, covering a small portion of the lower branch where MI can also be triggered (as shown in Fig. 3a in the main text). This is discussed in more detail in the next section.

Conversely, when the continuous-wave operating point (i.e. $\alpha$ and $\rho$) is fixed, a range of modes will experience positive modulational gain. The largest gain happens when $\zeta=\alpha-2|\rho|^2$, meaning that the cross-phase modulation compensated the phase mismatch to match the given detuning. This can happen for the regions to the red side of the $\alpha=2|\rho|^2$ line, where the required $\zeta$ is positive and $m$ number can be solved accordingly. To the blue side of the $\alpha=2|\rho|^2$ line the condition can not be satisfied, and making $\zeta$ smaller increases the gain. This would make the pump mode have the largest gain, but the gain is countered by injection locking, which keeps the continuous-wave power at the operating point. The neighboring modes with relative mode numbers of $\pm 1$ instead receive the largest gain.

Domains and domain walls form from the fluctuations of the continuous-wave solution on the unstable branch. The initial fluctuation is dominated by the mode with the largest gain, and divides the resonator into $m$ sections with a slightly higher power and $m$ sections with a slightly lower power compared to the equilibrium. Subsequent evolutions will create $m$ high- and low-field domains based in the initial field pattern in the resonator. Therefore operating points located to the blue side of the $\alpha=2|\rho|^2$ line initiates single-pulse formations, while operating points to the red side initiates multiple-pulse formations, in which case the pulse number can be estimated from $\zeta=\alpha-2|\rho|^2$. As noted in the main text, the exact pulse number is subject to domain wall collisions and other transient processes, and the pulse number calculated this way remains as an estimate. For the single pulse regime, as the gains on the modes with small $m$ numbers are similar, the final state also depends on the initial fluctuations in the equilibrium.

\section{Dark pulse creation in DI and Turing roll regimes}

\begin{figure*}
\includegraphics[width=170mm]{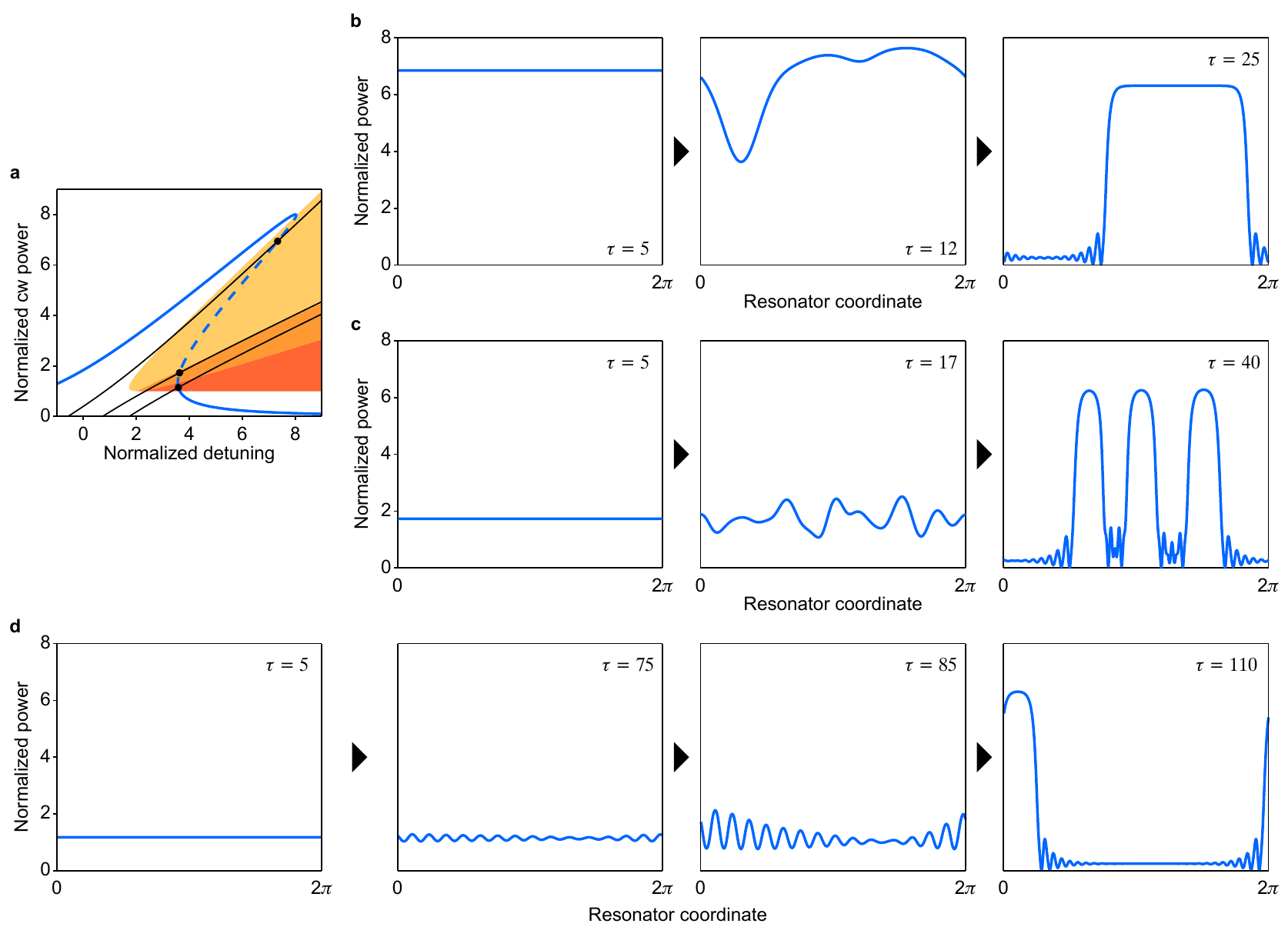}
\caption{\label{FIG8}
Comparison for pulse generation in different regimes.
(a) Resonator pumping curve for $|f|^2=8$ (blue) and laser locking curve for $\phi=-\pi/3$ (black, left), $\phi=2\pi/5$ (black, middle) and $\phi=2\pi/3$ (black, right). These three feedback phases place the continuous-wave operating point (black dots) in the single pulse, multiple pulses and Turing roll region, respectively. They are used for simulations resulting in panels (b)-(d), respectively.
(b) Simulated waveforms for $\phi=-\pi/3$.
First panel: At $\tau=5$, the system approaches the continuous-wave operating point.
Second panel: At $\tau=12$, fluctuations become visible and provide seeding for the dark pulse.
Third panel: At $\tau=25$, the system settles to a single-pulse state.
(c) Simulated waveforms for $\phi=2\pi/5$.
First panel: At $\tau=5$, the system approaches the continuous-wave operating point.
Second panel: At $\tau=17$, fluctuations become visible and provide seeding for the dark pulses.
Third panel: At $\tau=40$, the system settles to a multiple-pulse state.
(d) Simulated waveforms for $\phi=2\pi/3$.
First panel: At $\tau=5$, the system approaches the continuous-wave operating point.
Second panel: At $\tau=75$, Turing rolls become visible.
Third panel: At $\tau=85$, fluctuations with low $m$ appear on top of the Turing rolls and provides seeding for the dark pulse.
Fourth panel: At $\tau=110$, the system settles to a single dark pulse state.
}
\end{figure*}

If the continuous-wave operating point is located in the Turing roll regime, Turing rolls will start to form in the resonator. However, most of these rolls are unstable as the local power may exceed the unstable $\rho_\mathrm{M}$ branch and the field will be pushed towards $\rho_\mathrm{H}$ \cite{godey2014stability}. While for conventional resonators the system may converge to the $\rho_\mathrm{H}$ continuous-wave equilibrium, this is not possible for injection-locked system studied here as this would pull the system off the locking curve. Sidebands with low $m$ numbers are still amplified, and usually result in dark pulses in the way similar to generating pulses from DI as studied above. Examples of the different cases are compared in Fig. \ref{FIG8}.

\section{Duty cycle and feedback phase}

Here we study the dependence of the duty cycle on the feedback phase $\phi$. For simplicity, we ignore the width of the domain walls, and the field can be regarded as consisting of $w\%$ of high-field domain and $1-w\%$ of low-field domain. We can thus approximate $\rho=w\% \rho_\mathrm{H} +(1-w\%)\rho_\mathrm{L}$ and $P=w\% |\rho_\mathrm{H}|^2 +(1-w\%) |\rho_\mathrm{L}|^2$. We note that, as we have ignored the domain wall widths, these averages remain independent of the number of domains and the number of dark pulses. The duty cycle can thus be related to the feedback phase via the locking condition:
\begin{equation}
\mathrm{Im}\left[\frac{e^{i\phi}}{1+i\alpha-2iP}\frac{\rho}{f}\right]=0\ \ \rightarrow\ \ 
\phi=\mathrm{Arg}\left[\frac{1+i\alpha_\mathrm{MP}-2i\left(w\% |\rho_\mathrm{H}|^2 +(1-w\%) |\rho_\mathrm{L}|^2\right)}{w\% \rho_\mathrm{H} +(1-w\%)\rho_\mathrm{L}}\right]
\end{equation}
where $f$ is assumed to be real and $\alpha_\mathrm{MP}$ is the Maxwell Point detuning corresponding to $f$. The duty cycle can be solved numerically after the values of $\rho_\mathrm{L}$ and $\rho_\mathrm{H}$ are obtained at the Maxwell point.

In the cases where domain wall widths are non-negligible, for example when the $\beta_2$ is large, the internal structure of the domain wall needs to be considered to calculate the average field and power. The walls may also interact with each other if their tails overlap. In these cases the above equation provides an estimate of the duty cycle and numerical simulations should be used for a more accurate result. The finite domain wall width will also make the duty cycle dependent on the number of dark pulses.

\section{Additional simulation results}

\begin{figure}
\includegraphics[width=170mm]{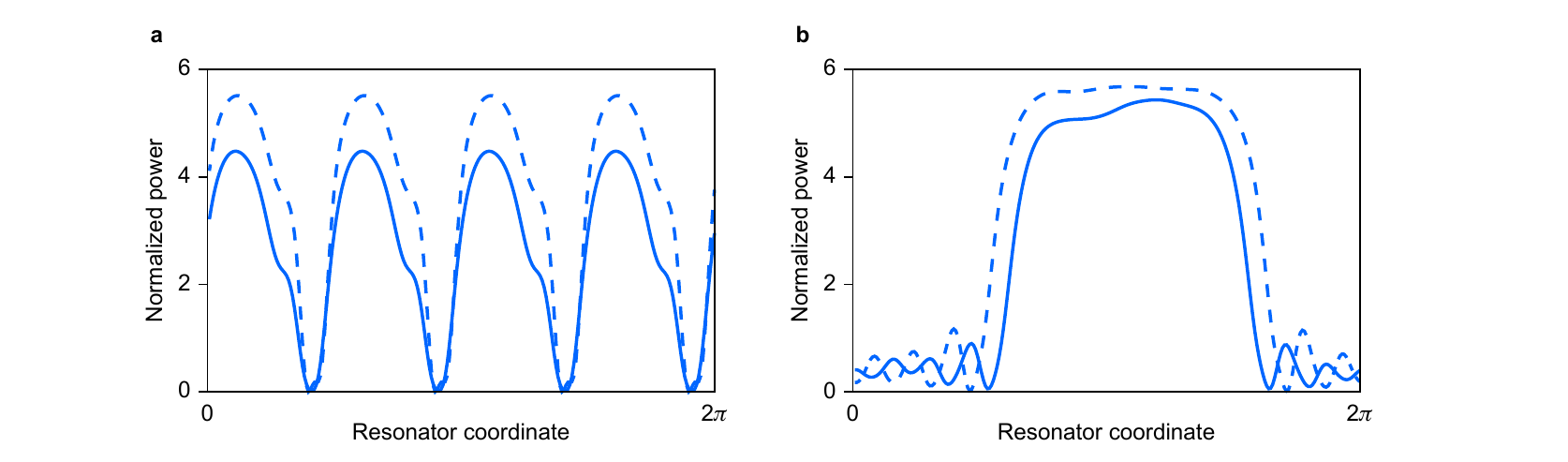}
\caption{\label{FIG9}
Numerically simulated waveforms with randomized backscattering on each mode. Pumping strength is taken as $|f|^2=8$. (a) A multiple-pulse state for $\phi=\pi/3$ (solid curve) and $\phi=-\pi/3$ (dashed curve). The duty cycles are 59\% and 72\%, respectively. (b) A single-pulse state for $\phi=\pi/3$ (solid curve) and $\phi=-\pi/3$ (dashed curve). The duty cycles are 45\% and 53\%, respectively.
}
\end{figure}

\begin{figure}
\includegraphics[width=170mm]{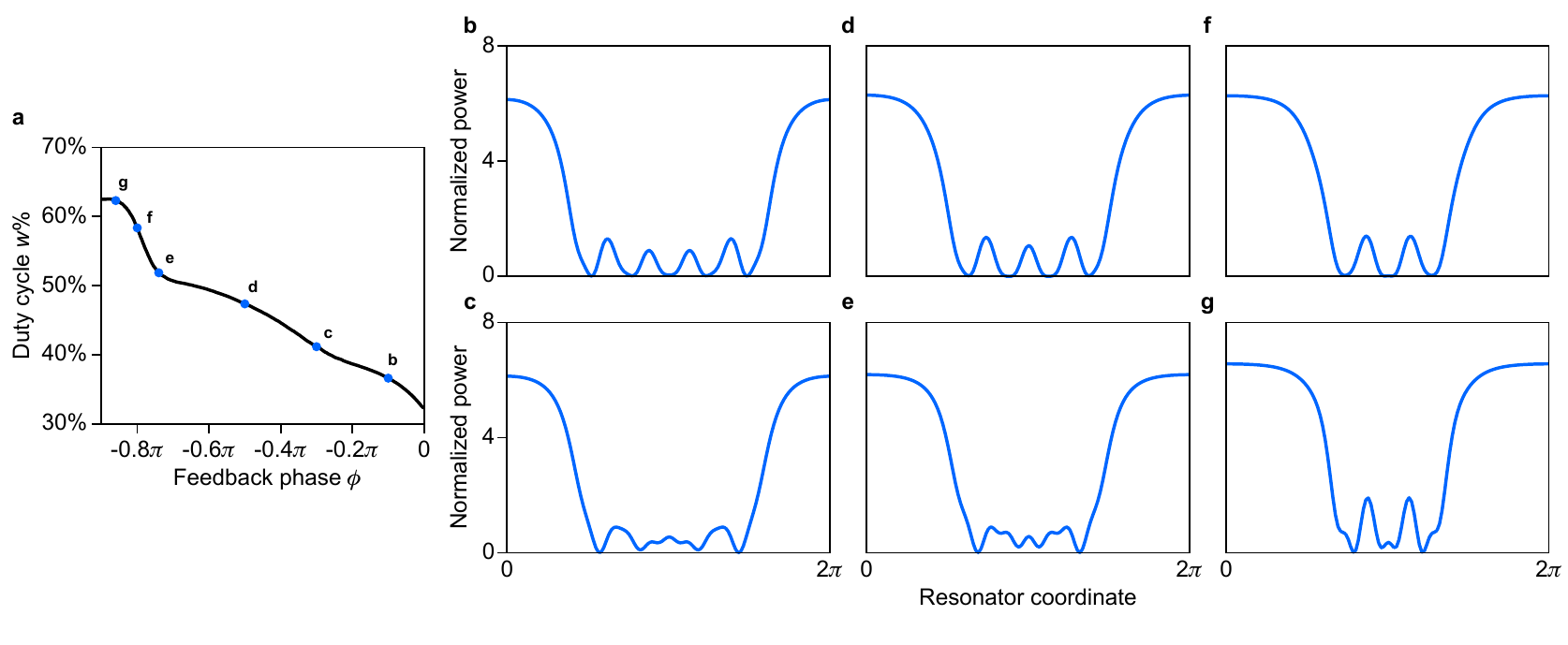}
\caption{\label{FIG10}
Numerically simulated waveforms with large modal dispersion ($\beta_2=0.16$). Pumping strength is taken as $|f|^2=8$. (a) Plot of simulated duty cycle versus feedback phase. The data is obtained by starting at $\phi=0$ and decreasing the phase adiabatically. (b)-(g) Intracavity power at the corresponding points in (a).
}
\end{figure}

It is noted in the main text that the domain walls and the associated dark pulses may have irregular shapes due to the distributed backscattering in the resonator. Numerical simulations have been performed with randomized backscattering on each mode and the simulated waveforms are shown in Fig. \ref{FIG9}. Backscattering may cause the pair of domain walls to become asymmetric and the domains to be weakly oscillating. Although the entire field circulates along the resonator in the lab frame, the backscattering does not average out as the propagating field profile are not homogeneous themselves. The spatial structures will be carried over when energy is transferred to and from the reflected field, which leads to distortions of the domains and domain walls. However, the increase in duty cycle with respect to decreasing feedback phase can still be observed, regardless of whether the underlying states consist of multiple pulses or a single pulse.

Large $\beta_2$ of the mode leads to wider domain wall widths, stronger domain wall interactions, and modifies the dependence of duty cycle on the feedback phase. Figure \ref{FIG10}a shows a typical dependence of duty cycle on feedback phase when $\beta_2$ is large. Undulations are present on the curve, and for certain range of feedback phase the slope of the curve becomes smaller. This is a result of alternating attractive or repulsive forces acting on the domain walls, and are closely related to the snaking bifurcations describing non-injection-locked dark pulses \cite{parra2016origin}. Waveforms shown in Figs. \ref{FIG10}b to \ref{FIG10}g further confirm that the interactions between domain wall tails leads to duty cycle changes. Other effects such as dispersive waves \cite{yi2017single} could also be present that changes the behavior of domain walls and may lead to even weaker dependence of duty cycle on the feedback phase.

\end{widetext}

\noindent\bibliography{main.bib}

\end{document}